\documentclass[pre,onecolumn,amsmath,amssymb]{revtex4}
\usepackage{graphicx}
\begin{document}

\title{Universal Critical Behavior of Noisy Coupled Oscillators: a
Renormalization Group Study}

\author{Thomas Risler$^{1,2}$, Jacques Prost$^{3,2}$ and Frank J\"ulicher$^{1}$}
\affiliation{$^1$Max-Planck-Institut f\"ur Physik komplexer Systeme,
N\"othnitzerstrasse 38, 01187 Dresden, Germany}
\affiliation{$^2$Physicochimie Curie (CNRS-UMR 168), Institut Curie,
26 rue d'Ulm, 75248 Paris Cedex 05, France} \affiliation{$^3$Ecole
Sup\'erieure de Physique et de Chimie Industrielles de la Ville de
Paris, 10 rue Vauquelin, 75231 Paris Cedex 05, France}

\date{\today}

\begin{abstract}
We show that the synchronization transition of a large number of
noisy coupled oscillators is an example for a dynamic critical point
far from thermodynamic equilibrium. The universal behaviors of such
critical oscillators, arranged on a lattice in a $d$-dimensional
space and coupled by nearest neighbors interactions, can be studied
using field theoretical methods. The field theory associated with
the critical point of a homogeneous oscillatory instability (or Hopf
bifurcation of coupled oscillators) is the complex Ginzburg-Landau
equation with additive noise. We perform a perturbative
renormalization group (RG) study in a $4-\epsilon$ dimensional
space. We develop an RG scheme that eliminates the phase and
frequency of the oscillations using a scale-dependent oscillating
reference frame. Within a Callan-Symanzik RG scheme to two-loop
order in perturbation theory, we find that the RG fixed point is
formally related to the one of the model $A$ dynamics of the real
Ginzburg-Landau theory with an $O(2)$ symmetry of the order
parameter. Therefore, the dominant critical exponents for coupled
oscillators are the same as for this equilibrium field theory. This
formal connection with an equilibrium critical point imposes a
relation between the correlation and response functions of coupled
oscillators in the critical regime. Since the system operates far
from thermodynamic equilibrium, a strong violation of the
fluctuation-dissipation relation occurs and is characterized by a
universal divergence of an effective temperature. The formal
relation between critical oscillators and equilibrium critical
points suggests that long-range phase order exists in critical
oscillators above two dimensions.
\end{abstract}
\pacs{.}

\maketitle

\section{\label{Se.Intro}Introduction}

Equilibrium systems consisting of a large number of degrees of
freedom exhibit phase transitions as a consequence of the collective
behavior of many components \cite{fish67,brez76c,hohe77}. The
universal behaviors near critical points have been studied
extensively using field theoretical methods and renormalization
group (RG) techniques
\cite{kada67,wils71a,wils71b,wils74,wils75,hohe77,domi78,kogu79,wils83,fish98,zinnB}.
In systems driven far from thermodynamic equilibrium, collective
behaviors can lead to dynamic instabilities and non-equilibrium
phase transitions \cite{schm95,fishDS98,hinr00,taub02}. While the
study of non-equilibrium critical points has remained a big
challenge, RG methods have in some cases been applied
\cite{medi89,frey94,frey94b,jans97,gold99}.

An important example for non-equilibrium critical behavior is a
homogeneous oscillatory instability or Hopf bifurcation of coupled
oscillators \cite{risl04}. Such instabilities are important in many
physical, chemical, and biological systems \cite{stroB,bergB}. From
the point of view of statistical physics, phase coherent
oscillations result as the collective behavior of a large number of
degrees of freedom in the thermodynamic limit. For a system of
finite size, fluctuations destroy the phase coherence of the
oscillations and the singular behaviors characteristic of a Hopf
bifurcation are concealed.

This general idea can be illustrated by individual oscillators
arranged on a lattice in a $d$-dimensional space and coupled to
their nearest neighbors. As a consequence of fluctuations in the
system, each oscillator is subject to a noise source. For small
coupling strength and as a result of fluctuations, the oscillators
have individual phases and exhibit a limited coherence time of
oscillations. In the thermodynamic limit, a global phase emerges
beyond a critical coupling strength where oscillations become
coherent over large distances. As this critical point is approached
from the disordered phase, a correlation length, which corresponds
to the characteristic size of the domains of synchronized
oscillators, diverges. At the same time, oscillations become
coherent over long periods of time and long range phase order
appears. The divergence of the correlation length allows a
description of the critical behaviors of spatially extended systems
in terms of continuous field theories and is characteristic of
scale-invariance in the critical regime. It constitutes the
foundation of the renormalization group (RG) theory, which explains
the emergence of universality at critical points
\cite{kada67,wils71a,wils71b,wils83,fish98,zinnB}.

In this paper, we perform an RG study of the critical behaviors of a
collection of oscillators, distributed on a $d$-dimensional lattice
and coupled via nearest neighbors interactions. We approach the
critical point from the disordered phase. Close to criticality, the
large scale properties of the array of oscillators are described by
a dynamic field theory that is given by the complex Ginzburg-Landau
equation with an additive noise term. We apply field theoretical
perturbation theory in a $d=4-\epsilon$ dimensional space and
introduce an RG scheme that is appropriate for the study of coupled
oscillators.

The outline of the paper is as follows: In Section \ref{Se.GenTheo},
we present the general field theoretical framework for the complex
Ginzburg-Landau equation. We introduce in Section \ref{Se.OscFrame}
an oscillating reference frame that is essential to define the RG
procedure for oscillating critical systems and in which the phase
and frequency of the oscillations are eliminated. The correlation
and response functions of critical oscillators in mean field theory
are discussed in Section \ref{Se.MF}. These mean field results are
relevant above the critical dimension $d_c=4$. For $d<4$, mean field
theory breaks down. We discuss in Section \ref{Se.RGWils} the
renormalization group of the complex Ginzburg-Landau field theory
using a Wilson's RG scheme for which the renormalization procedure
for critical oscillators can be introduced most clearly. One-loop
order calculations in perturbation theory are presented, but further
calculations are necessary to characterize the correct qualitative
structure of the RG flow. In Section \ref{Se.RG}, we present the
Callan-Symanzik's RG scheme for systems of coupled oscillators and
calculate its beta functions as well as its complete RG flow and
fixed points to two-loop order in perturbation theory. The physical
properties of the oscillating system are characterized by
correlation and response functions. In Section
\ref{Se.AsymCritBehav}, we discuss the asymptotic behaviors of these
functions in the critical regime using the RG flow and applying a
matching procedure. The formal relation of the RG fixed point for
critical oscillators to the fixed point of a real Ginzburg-Landau
theory (which satisfies a fluctuation-dissipation (FD) relation)
leads to an emergent symmetry at the critical point. However, the
system operates far from thermodynamic equilibrium and breaks the FD
relation. The degree of this violation can be characterized by the
introduction of a frequency-dependent effective temperature which
diverges with a universal anomalous power-law at the critical point.
We conclude our presentation with a discussion of the general
properties of critical oscillators, their relations to equilibrium
critical points and possible experimental systems for which the
critical behaviors discussed here could be observed in the future.

\section{\label{Se.GenTheo}Field theory of coupled oscillators}

\subsection{\label{Sse.FieldTheo}Complex Ginzburg-Landau field theory}

The generic behavior of a nonlinear oscillator in the vicinity of a
Hopf bifurcation can be described by a dynamic equation for a
complex variable $Z$ characterizing the phase and amplitude of the
oscillations \cite{stroB}. This variable can be chosen such that its
real part is, to linear order, related to a physical observable,
e.g. the displacement $X(t)$ generated by a mechanical oscillator:
$X(t)={\rm Re}(Z(t))+$nonlinear terms. In the presence of a periodic
stimulus force $F(t)=\tilde F e^{-i\omega t}$ with a frequency
$\omega$ close to the oscillation frequency $\omega_0$ at the
bifurcation, the generic dynamics obeys
\begin{equation}\label{HopfNF}
\partial_{t}Z=-(r+i\omega_{0})Z-(u+iu_a)\left| Z \right|^{2}Z+\Lambda^{-1} e^{i\theta} F(t).
\end{equation}
For $F=0$ and $r>0$, the static state $Z=0$ is stable. The system
undergoes a Hopf bifurcation at $r=0$ and exhibits spontaneous
oscillations for $r<0$. The nonlinear term, characterized by the
coefficients $u$ and $u_a$, stabilizes the oscillation amplitude for
$u>0$. The external stimulus appears linearly in this equation and
couples in general with a phase shift $\theta$ \cite{cama00,egui00}.
In the case of a mechanical oscillator, the coefficient $\Lambda$
has units of a friction.

Coupling many oscillators in a field theoretic continuum limit leads
to the complex Ginzburg-Landau equation \cite{newe71,aran02} with
additive noise and external forcing terms:
\begin{equation}\label{cgle}
\partial_{t}Z=-(r+i\omega_{0})Z+(c+ic_{a})\Delta Z-(u+iu_{a})\left| Z \right|^{2}Z+\Lambda^{-1} e^{i\theta} F+\eta.
\end{equation}
Here, the complex variable $Z(\mathbf{x},t)$ becomes a field defined
at positions $\mathbf{x}$ in a $d$-dimensional space and $\Delta$
denotes the Laplace operator in this space. The coefficients $c$ and
$c_a$ characterize the local coupling of oscillators and the effects
of fluctuations are described via a complex random forcing term
$\eta(\mathbf{x},t)$, which will be chosen Gaussian with zero mean
value, i.e. $\langle\eta(\mathbf{x},t)\rangle=0$. As far as long
time and long wavelength properties are concerned, the correlation
times of the noise can be neglected and white noise can be used.

For a vanishing external field $F(\mathbf{x},t)$ and in the absence
of fluctuations, Eq. (\ref{cgle}) is invariant with respect to phase
changes of the oscillations:
\begin{equation}
Z\rightarrow Ze^{i\phi}.
\end{equation}
This symmetry reflects the fact that only phase-invariant terms
contribute to the dominant critical behaviors studied here. Indeed,
the Hopf bifurcation is associated with the emergence of a non-zero
oscillatory mode, which dominates the critical behaviors and for
which time-translational invariance and phase invariance are
equivalent. The noise correlations can therefore be chosen such that
they respect phase invariance in the problem:
\begin{eqnarray}
\langle\eta(\mathbf{x},t)\eta(\mathbf{x}',t')\rangle&=&0\nonumber\\
\langle\eta(\mathbf{x},t)\eta^*(\mathbf{x}',t')\rangle&=&4D\delta^{d}(\mathbf{x}-\mathbf{x}')\delta(t-t').
\end{eqnarray}
Here $D$ is a real and positive coefficient characterizing the
amplitude of the noise, and $\delta$ and $\delta^{d}$ represent
Dirac distributions respectively in $1$ and $d$ dimensions.

\subsection{Physical correlation and response functions}

Since the physical variables of interest are real, we decompose the
complex fields $Z$ and $F$ into their real and imaginary parts by
$Z=\psi_1+i\psi_2$ and $F=F_1+iF_2$. We focus on two important
functions which characterize the behavior of the system, namely the
two-point auto-correlation function $C_{\alpha\beta}$ and the linear
response function $\chi_{\alpha\beta}$ to an applied external
forcing term. They are defined as
\begin{eqnarray}
&&C_{\alpha\beta}(\mathbf{x}-\mathbf{x}',t-t')
=\langle\psi_{\alpha}(\mathbf{x},t)\psi_\beta(\mathbf{x}',t')\rangle_c\nonumber\\
&&\langle \psi_\alpha(\mathbf{x},t)\rangle=\int d^d x'\,d
t'\,\chi_{\alpha\beta}(\mathbf{x}-\mathbf{x}',t-t')
F_\beta(\mathbf{x}',t')+O(\vert F\vert^2),
\end{eqnarray}
where $\langle...\rangle_c$ denotes  a connected correlation
function \footnote{Note that because of space-time translational
invariance of the theory, these functions depend only on the
differences $\mathbf{x}-\mathbf{x}'$ and $t-t'$.}. Because of phase
invariance, these functions obey the following symmetry relations:
\begin{equation}\label{RotSym}
C_{11}=C_{22} \qquad \textrm{and} \qquad C_{21}=-C_{12},
\end{equation}
with similar relations for the linear response function $\chi_{\alpha\beta}$.

\subsection{\label{Sse.F.TheoFormNue} Field theoretical representation}

The correlation and response functions $C_{\alpha\beta}$ and
$\chi_{\alpha\beta}$ can be conveniently expressed using a
field-theoretical formalism. We introduce the Martin-Siggia-Rose
response field $\tilde{\psi}_{\alpha}$ \cite{mart73} and apply the
Janssen-De Dominicis formalism \cite{jans76,domi76} to write the
following generating functional:
\begin{equation}\label{FctGen1}
Z_P\left[\tilde{I}_{\alpha},I_{\alpha}\right]=\int\mathcal{D}\left[\psi_{\alpha}\right]
\,\mathcal{D}\left[-i\tilde{\psi}_{\alpha}\right]\,\exp\left\{\mathcal{S}_P\left[\tilde{\psi}_{\alpha},\psi_{\alpha}\right]+\int
d^d x\,d t\,
\left[\tilde{I}_{\alpha}\,\tilde{\psi}_{\alpha}+I_{\alpha}\psi_{\alpha}\right]\right\},
\end{equation}
where the ``physical'' action $\mathcal{S}_P$ is given
by\footnote{Note that we do not write the Jacobian term that appears
in general in this formalism. Its role is indeed to compensate loops
$\langle\psi_{\alpha}\tilde{\psi}_{\beta}\rangle_0$ that begin and
end at the same vertex in a perturbative development. This choice,
followed by the prescription that such self-loops vanish, is
consistent and preserves causality \cite{zinnB}.}
\begin{equation}\label{Act1}
\mathcal{S}_P\left[\tilde{\psi}_{\alpha},\psi_{\alpha}\right]=\int
d^d x\,d t\, \left\{
D\tilde{\psi}_{\alpha}\tilde{\psi}_{\alpha}-\tilde{\psi}_{\alpha}\left[\partial_{t}\psi_{\alpha}+(R_{\alpha\beta}+\omega_0
\,\varepsilon_{\alpha\beta})\psi_{\beta}+U_{\alpha\beta}\psi_{\beta}\psi_{\gamma}\psi_{\gamma}\right]\right\}.
\end{equation}
Here
$R_{\alpha\beta}=(r-c\Delta)\delta_{\alpha\beta}-c_{a}\Delta\,\varepsilon_{\alpha\beta}$
and $U_{\alpha\beta}=u \delta_{\alpha\beta} + u_a
\varepsilon_{\alpha\beta}$, with
$\varepsilon_{21}=-\varepsilon_{12}=1$ and
$\varepsilon_{11}=\varepsilon_{22}=0$. The field $\tilde I_\alpha$
is related to the external force $F_{\alpha}$ in the equation of
motion (Eq. (\ref{cgle})) by
\begin{equation}
\tilde I_\alpha=\Lambda^{-1}\Omega_{\alpha\beta}(\theta)F_\beta,
\end{equation}
where
$\Omega_{\alpha\beta}(\theta)$ denotes the rotation matrix by an angle
$\theta$ in two dimensions:
\begin{equation}\label{RotMat2.2}
\Omega_{\alpha\beta}(\theta)=\left( \begin{array}{cc}
\cos(\theta) & -\sin(\theta)\\
\sin(\theta) & \cos(\theta)
\end{array} \right).
\end{equation}
Correlation and response functions are given by derivatives of the
generating functional. We have
\begin{eqnarray}
C_{\alpha\beta}({\bf x}-{\bf x}',t-t')&=&\frac{\delta^2\ln Z}
{\delta I_\alpha({\bf x},t)\delta I_\beta({\bf x}',t')}\vert_{I_\alpha,\tilde I_\alpha=0}\\
\chi_{\alpha\beta}({\bf x}-{\bf
x}',t-t')&=&\Lambda^{-1}\Omega_{\gamma\beta}(\theta)\frac{\delta^2\ln
Z}{\delta I_\alpha({\bf x},t)\delta \tilde I_\gamma({\bf x}',t')}
\vert_{I_\alpha,\tilde I_\alpha=0}.
\end{eqnarray}

\section{\label{Se.OscFrame}Field theory in an oscillating reference frame}

\subsection{\label{Sse.AmpliEq}Amplitude equation}

The frequency $\omega_0$ and the phase $\theta$ can be eliminated
from Eq. (\ref{cgle}) by a time-dependent variable transformation of
the form $Y\equiv e^{i\omega_0t}Z$, $H\equiv
e^{i\omega_0t}\Lambda^{-1} e^{i\theta}F$ and $\zeta\equiv
e^{i\omega_0t}\eta$, where $Y$ denotes the oscillation amplitude,
$H$ is a forcing amplitude and $\zeta$ is a transformed noise that
has the same correlators as $\eta$. This procedure leads to the
amplitude equation
\begin{equation}\label{cgley}
\partial_{t}Y=-rY+(c+ic_{a})\Delta Y-(u+iu_{a})\left| Y \right| ^{2}Y+H+\zeta.
\end{equation}
Defining two real fields $\phi_\alpha$ by $Y=\phi_{1}+i\phi_{2}$, Eq. (\ref{cgley}) reads
\begin{equation}\label{MatEvol}
\partial_{t}\phi_{\alpha}=-R_{\alpha\beta}\phi_{\beta}-U_{\alpha\beta}\phi_{\beta}\phi_{\gamma}\phi_{\gamma}+H_{\alpha}+\zeta_{\alpha},
\end{equation}
where $H=H_1+iH_2$.

The correlation and response functions
$G_{\alpha\beta}=\langle\phi_{\alpha}\phi_{\beta}\rangle_c$ and
$\gamma_{\alpha\beta}=\langle\phi_{\alpha}\tilde{\phi}_{\beta}\rangle_c$
of the fields $\phi_\alpha$ are related to the physical correlation
and response functions by
\begin{eqnarray}\label{Phys/MathCoRespFct}
C_{\alpha\beta}(\mathbf{x},t)&=&\Omega_{\alpha\sigma}(-\omega_0\,t)\,G_{\sigma\beta}(\mathbf{x},t)\nonumber\\
\chi_{\alpha\beta}(\mathbf{x},t)&=&\Lambda^{-1}\,\Omega_{\alpha\sigma}(\theta-\omega_0\,t)\,\gamma_{\sigma\beta}(\mathbf{x},t).
\end{eqnarray}
Similar time-dependent transformations exist between higher order
correlation and response functions of the physical fields and those
calculated in the oscillating reference frame.

\subsection{\label{Sse.Anal}Analogy with an equilibrium critical point in a particular case}

For the particular case where $c_a=0$ and $u_a=0$, Eq.
(\ref{MatEvol}) becomes identical to  the model A dynamics of a real
Ginzburg-Landau field theory with an $O(2)$ symmetry of the order
parameter \cite{hohe77}. The critical behavior of this theory at
thermodynamic equilibrium has been extensively studied
\cite{brez74,baus76,taub92}. This leads, in this particular case, to
a formal analogy between an equilibrium phase transition and a Hopf
bifurcation. The amplitude $Y$ plays the role of the order parameter
of the transition. The disordered phase with $\langle Y\rangle=0$
corresponds to noisy oscillators that are not in synchrony, while
nonzero order $\langle Y\rangle$ implies the existence of a global
phase and amplitude of synchronous oscillations with frequency
$\omega_0$. The correlation lengths  and times of the equilibrium
field theory correspond to lengths and times over which oscillators
are in synchrony. The correlation and response functions
$C_{\alpha\beta}$ and $\chi_{\alpha\beta}$ for this case can be
obtained from those of the equilibrium field theory by using Eq.
(\ref{Phys/MathCoRespFct}). Since the $O(2)$ symmetric theory at
thermodynamic equilibrium obeys an FD relation, a generic relation
between the correlation and response functions $C_{\alpha\beta}$ and
$\chi_{\alpha\beta}$ appears.

\subsection{\label{Sse.PerTheo}Generating functional}

The functions $G_{\alpha\beta}$ and $\gamma_{\alpha\beta}$, as well
as higher order correlation and response functions, can be formally
calculated using field theoretical techniques  (see e.g.
\cite{brez76c,zinnB}). The generating functional of the theory is
given by
\begin{equation}\label{FctGenPhi}
Z\left[\tilde{J}_{\alpha},J_{\alpha}\right]=\int\mathcal{D}\left[\phi_{\alpha}\right]
\,\mathcal{D}\left[-i\tilde{\phi}_{\alpha}\right]
\,\exp\left\{\mathcal{S}\left[\tilde{\phi}_{\alpha},\phi_{\alpha}\right]+\int
d^d x\,d t\,
\left[\tilde{J}_{\alpha}\tilde{\phi}_{\alpha}+J_{\alpha}\phi_{\alpha}\right]\right\},
\end{equation}
where we have introduced the Martin-Siggia-Rose response field
$\tilde{\phi}_{\alpha}$ \cite{mart73}. The associated action reads:
\begin{equation}\label{ActPhi}
\mathcal{S}\left[\tilde{\phi}_{\alpha},\phi_{\alpha}\right]=\int
d^{d} x\,d t\,\left\{D\tilde{\phi}_{\alpha}\tilde{\phi}_{\alpha}
-\tilde{\phi}_{\alpha}\left[\partial_{t}\phi_{\alpha}+R_{\alpha\beta}\phi_{\beta}\right]
-U_{\alpha\beta}\tilde{\phi}_{\alpha}\phi_{\beta}\phi_{\gamma}\phi_{\gamma}\right\}.
\end{equation}
Correlation and response functions are given by
\begin{eqnarray}
G_{\alpha\beta}({\bf x}-{\bf x}',t-t')&=&\frac{\delta^2\ln Z}
{\delta J_\alpha({\bf x},t)\delta J_\beta({\bf x}',t')}\vert_{J_\alpha,\tilde J_\alpha=0}\\
\gamma_{\alpha\beta}({\bf x}-{\bf x}',t-t')&=&\frac{\delta^2\ln
Z}{\delta J_\alpha({\bf x},t)\delta \tilde J_\beta({\bf x}',t')}
\vert_{J_\alpha,\tilde J_\alpha=0}.
\end{eqnarray}
The effective action of the theory is defined by
\begin{equation} \label{EffActPhi}
\Gamma\left[\tilde{\Phi}_{\alpha},\Phi_{\alpha}\right]=\int d^d x
\,d t\, \left[\tilde{J}_{\alpha}\tilde{\Phi}_{\alpha}
+J_{\alpha}\Phi_{\alpha}\right]-\ln{Z\left[\tilde{J}_{\alpha},J_{\alpha}\right]},
\end{equation}
with
\begin{equation}\label{EffActPhi.1}
\Phi_{\alpha}(\mathbf{x},t)=\frac{\delta\ln{Z}}{\delta
J_{\alpha}(\mathbf{x},t)} \quad \textrm{and} \quad
\tilde{\Phi}_{\alpha}(\mathbf{x},t)=\frac{\delta\ln{Z}}{\delta
\tilde{J}_{\alpha}(\mathbf{x},t)}.
\end{equation}

In order to perform calculations perturbatively, we split the action
into a harmonic or ``Gaussian'' part, and a quartic or
``interaction'' part as:
$\mathcal{S}=\mathcal{S}_0+\mathcal{S}_{\rm{int}}$. $\mathcal{S}_0$
is given by
\begin{eqnarray}\label{FreeAct}
\mathcal{S}_0&=&\int_{k}
\left\{\tilde{\phi}_{\alpha}(k)\left(
D\tilde{\phi}_{\alpha}(-k)-[i\omega\delta_{\alpha\beta}+R_{\alpha\beta}(-k)]\phi_{\beta}(-k)\right)\right\}\nonumber\\
&=&-\frac{1}{2}\int_{k} \underline{\phi}^t_{\alpha}(k)
\underline{\underline{A}}_{\alpha\beta}(-k)
\underline{\phi}_{\beta}(-k),
\end{eqnarray}
where
\begin{equation}
\underline{\phi}_{\alpha}=\left( \begin{array}{c}
\tilde{\phi}_{\alpha}\\
\phi_{\alpha}
\end{array}\right), \quad
\underline{\phi}^t_{\alpha}=\left(\tilde{\phi}_{\alpha},\phi_{\alpha}\right),
\end{equation}
\begin{equation}\label{AMat}
\underline{\underline{A}}_{\alpha\beta}(-k)=
\left( \begin{array}{cc}
-2D\delta_{\alpha\beta} & i\omega\delta_{\alpha\beta}+R_{\alpha\beta}(-k)\\
-i\omega\delta_{\beta\alpha}+R_{\beta\alpha}(k) & 0
\end{array} \right),
\end{equation}
and
$R_{\alpha\beta}(k)=(r+c\mathbf{q}^2)\delta_{\alpha\beta}+c_a\mathbf{q}^2\varepsilon_{\alpha\beta}$.
In these expressions and in the following, we label
$k=(\mathbf{q},\omega)$, $\int_{k}=\int_{\mathbf{q},\omega}$ denotes
$\int \frac{d^d q}{(2\pi)^d} \frac{d \omega}{2\pi}$, and we use the
following convention for Fourier-transforms:
\begin{equation}
f(\mathbf{x},t)=\int_{\mathbf{q},\omega}f(\mathbf{q},\omega)e^{i(\mathbf{q}.\mathbf{x}-\omega
t)}=\int_{k}f(k)e^{ik.x}.
\end{equation}
The interaction term of the action takes the form
\begin{equation}
\mathcal{S}_{\rm{int}}=-\int_{\{k_i\}}U_{\alpha\beta}\,\tilde{\phi}_{\alpha}(k_1)\phi_{\beta}(k_2)\phi_{\gamma}(k_3)\phi_{\gamma}(k_4)
\times(2\pi)^{d+1}\delta^{(d+1)}\left(\sum_{i}k_i\right).
\end{equation}
Graphic representations of the basic diagrams of the perturbation
theory are given in Appendix \ref{Ap.GraphRepr}.

\section{\label{Se.MF}Mean field theory}

Dimensional analysis reveals that for $d>4$ mean field theory
applies. In this case, the mean field approximation allows us to
calculate valid asymptotic expressions for the effective action of
the theory, as well as the two-point correlation and response
functions. In the framework of the Janssen-De Dominicis formalism
for dynamic field theoretical models, this approximation consists in
substituting the saddle-node value of the path-integral
(\ref{FctGenPhi}) to the full functional generator $Z$. We obtain
\begin{equation}\label{MFFctGene.1}
Z^{\rm
mf}\left[\tilde{J}_{\alpha},J_{\alpha}\right]=\exp\left\{\mathcal{S}\left[\tilde{\phi}^{\rm
mf}_{\alpha},\phi^{\rm mf}_{\alpha}\right]+\int d^d x\,d t\,
[\tilde{J}_{\alpha}\tilde{\phi}^{\rm
mf}_{\alpha}+J_{\alpha}\phi^{\rm mf}_{\alpha}]\right\},
\end{equation}
where $\mathcal{S}$ is given by (\ref{ActPhi}) and $\phi^{\rm
mf}_{\alpha}$ and $\tilde{\phi}^{\rm mf}_{\alpha}$ satisfy the
stationarity conditions at the saddle node
\begin{equation}
\frac{\delta\mathcal{S}}{\delta \tilde{\phi}_{\alpha}}= -\tilde J_\alpha \quad
\textrm{and} \quad \frac{\delta\mathcal{S}}{\delta \phi_{\alpha}}=
-J_\alpha,
\end{equation}
which give the mean field dynamic equations
\begin{eqnarray}\label{MFEq.1}
J_{\alpha}&=&-\partial_t\tilde{\phi}^{\rm
mf}_{\alpha}+R_{\beta\alpha}\tilde{\phi}^{\rm mf}_{\beta}
+U_{\beta\alpha}\tilde{\phi}^{\rm mf}_{\beta}\phi^{\rm
mf}_{\gamma}\phi^{\rm mf}_{\gamma} +2U_{\beta\sigma}\phi^{\rm
mf}_{\sigma}\phi^{\rm mf}_{\alpha}\tilde{\phi}^{\rm
mf}_{\beta}\nonumber\\
\tilde J_\alpha&=&-2D\tilde{\phi}^{\rm
mf}_{\alpha}+\left[\partial_t\phi_{\alpha}^{\rm mf}
+R_{\alpha\beta}\phi_{\beta}^{\rm
mf}+U_{\alpha\beta}\phi_{\beta}^{\rm mf}\phi_{\gamma}^{\rm
mf}\phi_{\gamma}^{\rm mf}\right].
\end{eqnarray}
The correlation and response functions can be obtained most easily
by first eliminating the nonphysical field $\tilde \phi^{\rm
mf}_\alpha$ and writing a mean field equation for $\phi^{\rm
mf}_\alpha$ only:
\begin{eqnarray}\label{MFEq.2}
2DJ_{\alpha}&=&\left[-\partial_t\delta_{\alpha\beta}+R_{\beta\alpha}+U_{\beta\alpha}\phi_{\gamma}^{\rm
mf}\phi_{\gamma}^{\rm mf}+2U_{\beta\sigma}\phi_{\sigma}^{\rm
mf}\phi_{\alpha}^{\rm mf}\right]\nonumber\\
&&\times\left[
\partial_t\phi_{\beta}^{\rm mf}+R_{\beta\sigma}\phi_{\sigma}^{\rm mf}+U_{\beta\sigma}\phi_{\sigma}^{\rm mf}\phi_{\gamma}^{\rm mf}\phi_{\gamma}^{\rm
mf}-\tilde{J}_{\beta}\right].
\end{eqnarray}
The mean field generating functional for the field $\phi^{\rm
mf}_\alpha$ obeying Eq. (\ref{MFEq.2}) is obtained from
(\ref{MFFctGene.1}) by eliminating the field $\tilde \phi^{\rm
mf}_\alpha$. It reads:
\begin{equation}\label{MFFctGene.2}
Z^{\rm
mf}\left[\tilde{J}_{\alpha},J_{\alpha}\right]=\exp\left\{-\frac{1}{4D}\int
d^d x\,d t\, \left[\left(\partial_{t}\phi_{\alpha}^{\rm
mf}-F_{\alpha}\left[\phi^{\rm
mf}_{\beta}\right]-\tilde{J}_{\alpha}\right)^2 +
J_{\alpha}\phi_{\alpha}^{\rm mf}\right]\right\},
\end{equation}
where
\begin{equation}
F_{\alpha}[\phi_{\beta}]=-R_{\alpha\beta}\phi_{\beta}-U_{\alpha\beta}\phi_{\beta}\phi_{\gamma}\phi_{\gamma}.
\end{equation}
Note that for $J_{\alpha}=0$ the stationary condition of the
generating functional (\ref{MFFctGene.2}) leads to
\begin{equation}\label{Eq.MFMatEvol}
\partial_{t}\phi^{\rm mf}_{\alpha}=-R_{\alpha\beta}\phi^{\rm mf}_{\beta}-U_{\alpha\beta}\phi^{\rm mf}_{\beta}\phi^{\rm mf}_{\gamma}\phi^{\rm
mf}_{\gamma}+\tilde{J}_{\alpha},
\end{equation}
which is compatible with Eq. (\ref{MFEq.2}).

The linear response and correlation functions are obtained as
\begin{eqnarray}
\gamma^{\rm
mf}_{\alpha\beta}(\mathbf{x}-\mathbf{x}';t-t')&=&\frac{\delta\phi^{\rm
mf}_{\alpha}(\mathbf{x},t)}{\delta \tilde
J_{\beta}(\mathbf{x}',t')}\vert_{J_{\alpha},\tilde{J}_{\alpha}\equiv
0}\nonumber\\
G^{\rm
mf}_{\alpha\beta}(\mathbf{x}-\mathbf{x}';t-t')&=&\frac{\delta\phi^{\rm
mf}_{\alpha}(\mathbf{x},t)}{\delta
J_{\beta}(\mathbf{x}',t')}\vert_{J_{\alpha},\tilde{J}_{\alpha}\equiv
0}.
\end{eqnarray}
Calculating these functions and applying the time-dependent
transformations (\ref{Phys/MathCoRespFct}), we obtain the physical
correlation and response functions in mean field theory
\begin{eqnarray}\label{MFCoResp}
\chi^{\rm{mf}}_{\alpha\beta}(\mathbf{q},\omega)&=&\frac{1}{\Lambda\Delta}
\left\{\left[(-i\omega+R)\cos{\theta}+\Omega_0\sin{\theta}\right]\delta_{\alpha\beta}
+\left[(-i\omega+R)\sin{\theta}-\Omega_0\cos{\theta}\right]\varepsilon_{\alpha\beta}\right\}\nonumber\\
C^{\rm{mf}}_{\alpha\beta}(\mathbf{q},\omega)&=&\frac{2D}{{\left| \Delta \right|}^2}
\left( \begin{array}{cc}
\omega^{2} + \Omega_{0}^2 + R^{2} & 2i\omega \Omega_0\\
-2i\omega \Omega_0 & \omega^{2} + \Omega_{0}^2 + R^{2}
\end{array} \right),
\end{eqnarray}
where $R=r+c\mathbf{q}^2$, $\Omega_{0}=\omega_{0}+c_{a}\mathbf{q}^2$
and $\Delta=(-i\omega+R)^{2}+\Omega_{0}^{2}$. The diagonal elements
of these matrices are given by
\begin{eqnarray}\label{MFCoRespDiag}
\chi_{11}^{\rm mf}(\mathbf{q},\omega)&=&\frac{1}{2\Lambda}\left[
\frac{e^{i\theta}}{R-i(\omega-\Omega_0)}+\frac{e^{-i\theta}}{R-i(\omega+\Omega_0)}\right]\nonumber\\
C_{11}^{\rm
mf}(\mathbf{q},\omega)&=&\frac{D}{R^2+(\omega-\Omega_0)^2}+\frac{D}{R^2+(\omega+\Omega_0)^2},
\end{eqnarray}
and the non-diagonal elements by
\begin{eqnarray}\label{MFCoRespNonDiag}
\chi_{12}^{\rm mf}(\mathbf{q},\omega)&=&\frac{i}{2\Lambda}\left[
\frac{e^{i\theta}}{R-i(\omega-\Omega_0)}-\frac{e^{-i\theta}}{R-i(\omega+\Omega_0)}\right]\nonumber\\
C_{12}^{\rm
mf}(\mathbf{q},\omega)&=&\frac{iD}{R^2+(\omega-\Omega_0)^2}-\frac{iD}{R^2+(\omega+\Omega_0)^2}.
\end{eqnarray}

Note finally that for $r=0$ and for the critical mode
$(\mathbf{q}=\mathbf{0},\omega=0)$, the response function of the
system is nonlinear even at small amplitudes. We have indeed in this
case
\begin{equation}
\delta\langle\left|Y(\mathbf{0},0)\right|\rangle^{\rm mf}\propto
\left| H\right|^{1/3}.
\end{equation}

\section{\label{Se.RGWils} Wilson's renormalization scheme}

For $d<4$, mean field theory breaks down and another approach is
necessary to investigate the critical behaviors of the theory. We
apply perturbative renormalization group (RG) methods using an
$\epsilon$ expansion near the upper critical dimension $d_c=4$
($d=4-\epsilon$) \cite{wils72,wils73,wils74}. We present here the RG
structure of the theory within a Wilson's momentum shell RG scheme
adapted to the renormalization of the complex Ginzburg-Landau field
theory. This scheme has the advantage to be conceptually transparent
and to provide a clear physical interpretation to the calculations.
However, for calculations beyond one-loop order, this technique is
less suited than the Callan-Symanzik RG scheme. The adaptation of
the latter to the complex Ginzburg-Landau theory is presented in
Section \ref{Se.RG}.

\subsection{\label{Sse.RGWilsGen}Renormalized fields}

The renormalization procedure within the Wilson's scheme is
performed as follows. We start from a dynamic functional of the
theory with a small distance cut-off $\Lambda$ in the integrals over
wave vectors. This cutoff corresponds to an underlying lattice of
mesh size $a\simeq 2\pi/\Lambda$. We interpret this as a microscopic
theory with an action of the form (\ref{Act1}), and associated
quantities are labeled with an superscript ``$0$''. We calculate the
effective action in an oscillating reference frame at a given scale
$\Lambda/b$, where $b=e^l$ is a dilatation coefficient larger than
1. This reference frame is defined such that, described in terms of
effective or renormalized parameters, the effective action has the
structure (\ref{ActPhi}). The renormalized quantities that satisfy
this requirement can be expressed as
\begin{eqnarray}\label{RenQties}
\mathbf{q}&=&b\mathbf{q}^0\nonumber\\
\omega&=&b^2 Z_\omega(b)\omega^0\nonumber\\
\phi_{\alpha}(\mathbf{q},t)&=&b^{-\frac{d+2}{2}}\sqrt{Z_{\phi}(b)}Z_{\omega}(b)\Omega_{\alpha\beta}(\hat{\omega}_0(b)t)
\psi^0_{\beta}(\mathbf{q}^0,t^0)\nonumber\\
\tilde{\phi}_{\alpha}(\mathbf{q},t)&=&b^{-\frac{d-2}{2}}\sqrt{Z_{\tilde{\phi}}(b)}Z_{\omega}(b)\Omega_{\alpha\beta}(\theta(b)+
\hat{\omega}_0(b)t)\tilde{\psi}^0_{\beta}(\mathbf{q}^0,t^0).
\end{eqnarray}
Here we have introduced scale-dependent $Z$ factors for the
renormalization of the two dynamic fields ($Z_{\phi}(b)$ and
$Z_{\tilde{\phi}}(b)$) and for the frequencies ($Z_\omega(b)$). In
addition to the usual RG transformations and scale dilatation
$\mathbf{q}=b\mathbf{q}^{(0)}$, the complex Ginzburg-Landau theory
requires us to perform time-dependent transformations between bare
and renormalized fields. Indeed, the effective theory is described
in a reference frame that oscillates with effective frequency $\hat
\omega_0(b)$ and phase $\theta(b)$ relatively to the reference frame
of the bare theory. The definition of the fields $\phi_\alpha$ and
$\tilde\phi_\alpha$ takes this relative rotation into account by
terms involving the rotation matrix $\Omega_{\alpha\beta}$ (Eq.
(\ref{RotMat2.2})). The scale-dependent oscillating reference frame
represents a key element of the RG procedure for oscillating
systems. For the microscopic theory, $\hat\omega_0(1)=\omega_0^0$,
$\theta(1)=0$ and the fields $\psi^0_\alpha$ and
$\tilde\psi^0_\alpha$ coincide with the fields introduced in Section
II.

\subsection{Renormalization group flow}

In order to determine the behavior of the effective parameters under
renormalization, we determine their variations with respect to a
small changes of the dilatation $\delta b/b$ (or $\delta l$).
Integrating over the momentum shell of wave vectors in the interval
$[\Lambda/e^{\delta l},\Lambda]$, we obtain an effective action
that, before rescaling, reads:
\begin{eqnarray}\label{ActNr1.2}
\mathcal{S}^{(1)}\left[\tilde{\phi}_{\alpha},\phi_{\alpha}\right]&=&\int_{\mathbf{q}}^{\Lambda/e^{\delta
l}}\int_{\omega}\left\{ (D+\delta
D)\tilde{\phi}_{\alpha}\tilde{\phi}_{\alpha}-\tilde{\phi}_{\alpha}\left[i(1+\delta\lambda_1)\omega+(r+\delta
r)+(c+\delta
c)\mathbf{q}^2\right]\phi_{\alpha}\right.\nonumber\\
&&\left.-\tilde{\phi}_{\alpha}\left[i\delta\lambda_2\omega+\delta\hat{\omega}_0+(c_a+\delta
c_a)\mathbf{q}^2\right]\varepsilon_{\alpha\beta}\phi_{\beta}\right\}
- \int_{k_1k_2k_3}^{\Lambda/e^{\delta l}} (U_{\alpha\beta}+\delta
U_{\alpha\beta})\tilde{\phi}_{\alpha}\phi_{\beta}\phi_{\gamma}\phi_{\gamma}.
\nonumber\\
\end{eqnarray}
Here, we have introduced variations of the parameters under this
procedure, which can be calculated perturbatively from the
``one-particle irreducible'' (1-PI) diagrams of the theory. Other
terms are either forbidden by symmetry properties of the theory or
irrelevant in the infra-red (IR) limit. Two new terms appear, which
were absent from Eq. (\ref{ActPhi}) and which correspond to the
coefficients $\delta\lambda_2$ and $\delta\hat{\omega}_0$. They
reflect the renormalization of frequency and phase. These terms are
absorbed by time-dependent variable transformations of the fields:
\begin{eqnarray}
\phi^{\rm
i}_{\alpha}(\mathbf{q},t)&=&\Omega_{\alpha\beta}\left(\delta\hat{\omega}_0^{\rm
i}\,t\right)\phi_{\beta}(\mathbf{q},t)\nonumber\\
\tilde{\phi}^{\rm
i}_{\alpha}(\mathbf{q},t)&=&\Omega_{\alpha\beta}\left(\delta\hat{\omega}_0^{\rm
i}\,t+\delta\theta\right)\phi_{\beta}(\mathbf{q},t),
\end{eqnarray}
where
\begin{eqnarray}
\delta\theta&=&\arctan{\left[-\delta\lambda_2(1+\delta\lambda_1)^{-1}\right]}\nonumber\\
\delta\hat{\omega}_0^{\rm
i}&=&\delta\hat{\omega}_0\cos{\delta\theta}+(r+\delta
r)\sin{\delta\theta}.
\end{eqnarray}

Rewriting the effective action in terms of the new fields
$\phi^i_\alpha$ and $\tilde \phi^i_\alpha$ requires a redefinition
of  the changes of all parameters. The so defined parameter changes
are label by a superscript ``i'' in the following. We finally
rescale all lengths by $\mathbf{q}'=e^{\delta l}\mathbf{q}$, and
introduce three $Z$ factors at the scale $l+\delta l$, such that the
effective action retains its form (\ref{ActPhi}) under an RG step.
Furthermore, we impose that $D(l+\delta l)=D(l)$ and $c(l+\delta
l)=c(l)$, i.e. that the coefficients $c$ and $D$ remain constants
under renormalization. As a result, the variations of the parameters
under a small renormalization step are given by
\begin{eqnarray}\label{RenPar}
c_a(l+\delta l)&=&\left[c_a(l)+\delta c_a^{\rm{i}}\right](1+\delta
c^{\rm{i}}/c)^{-1}\nonumber\\
r(l+\delta l)&=&e^{2\delta l}\left[r(l)+\delta r^{\rm{i}}\right](1+\delta
c^{\rm{i}}/c)^{-1}\nonumber\\
U_{\alpha\beta}(l+\delta l)&=&e^{(4-d)\delta
l}\left[U_{\alpha\beta}(l)+\delta
U_{\alpha\beta}^{\rm{i}}\right]\,(1+\delta
c^{\rm{i}}/c)^{-2}(1+\delta
D^{\rm{i}}/D)(1+\delta\lambda^{\rm{i}})^{-1}\nonumber\\
Z_\omega(l+\delta l)&=&Z_\omega(l)(1+\delta
c^{\rm{i}}/c)^{-1}(1+\delta\lambda^{\rm{i}})\nonumber\\
Z_{\phi}(l+\delta l)&=&Z(l)(1+\delta c^{\rm{i}}/c)^{3}(1+\delta
D^{\rm{i}}/D)^{-1}(1+\delta\lambda^{\rm{i}})^{-1}\nonumber\\
Z_{\tilde{\phi}}(l+\delta l)&=&\tilde{Z}(l)(1+\delta
c^{\rm{i}}/c)(1+\delta D^{\rm{i}}/D)(1+\delta\lambda^{\rm{i}})^{-1},
\end{eqnarray}
where
\begin{eqnarray}\label{Eq.InterParam.2}
\delta D^{\rm{i}}&=&\delta D\nonumber\\
\delta\lambda^{\rm{i}}&=&(C-1)+\delta\lambda_1C-\delta\lambda_2S\nonumber\\
\delta r^{\rm{i}}&=&r(C-1)+\delta rC-(\omega_0+\delta\omega_0)S\nonumber\\
\delta c^{\rm{i}}&=&c(C-1)+\delta cC-(c_a+\delta c_a)S\nonumber\\
\delta c_a^{\rm{i}}&=&c_a(C-1)+\delta c_aC+(c+\delta c)S\nonumber\\
\delta
U_{\alpha\beta}^{\rm{i}}&=&U_{\alpha\sigma}(\Omega_{\sigma\beta}(\delta\theta)-I_{\sigma\beta})+\delta
U_{\alpha\sigma}\Omega_{\sigma\beta}(\delta\theta).
\end{eqnarray}
Here, $C$ and $S$ denote $\cos{\delta\theta}$ and
$\sin{\delta\theta}$, respectively. The evolution of the effective
frequency and phase that enter the transformations (\ref{RenQties})
are given by
\begin{eqnarray}\label{EffPhaseFreq}
\theta(l+\delta l)&=&\theta(l)+\delta\theta\nonumber\\
\hat{\omega}_0(l+\delta l)&=&e^{2\delta
l}\left[\hat{\omega}_0(l)+\delta\hat{\omega}_0^{\rm{i}}\right]
(1+\delta c^{\rm{i}}/c)^{-1}(1+\delta\lambda^{\rm{i}}).
\end{eqnarray}

\subsection{\label{Sse.CritAsBehavWils}Correlation and response functions}

Using the RG transformations of the effective parameters, as well
the correlation and response functions of the renormalized fields
$\phi_\alpha$ and $\tilde\phi_\alpha$, we can write expressions for
the physical correlation and response functions, where the
transformations (\ref{RenQties}) have been used:
\begin{eqnarray}\label{Eq.TfFD.2}
C_{\alpha\beta}(\mathbf{q},t)&=&b^2Z_{\phi}(b)^{-1}Z_\omega(b)^{-2}\,\Omega_{\alpha\sigma}(-\omega_0(b)\,t)
\,G^{\rm{R}}_{\sigma\beta}(b\mathbf{q}\, ,\, b^{-2}Z_\omega(b)^{-1}t)\nonumber\\
\chi_{\alpha\beta}(\mathbf{q},t)&=&\sqrt{Z_{\phi}(b)Z_{\tilde{\phi}}(b)}^{-1}Z_\omega(b)^{-2}
\,\Omega_{\alpha\sigma}(\theta(b)-\omega_0(b)\,t)\,\gamma^{\rm{R}}_{\sigma\beta}(b\mathbf{q},\,b^{-2}Z_\omega(b)^{-1}t).
\end{eqnarray}
Here, the superscript``$R$'' indicates that the functions have to be
calculated using the renormalized set of parameters. The
scale-dependent renormalized frequency is given by
\begin{equation}\label{Eq.Ome0EffWil}
\omega_0(b)=b^{-2}Z_{\omega}(b)^{-1}\,\hat{\omega}_0(b),
\end{equation}
and $\theta(b)$ describes the scale-dependent phase lag between
external forcing and response of the system.

At the fixed point of the RG, the theory is scale invariant and,
consequently, the $Z$ factors exhibit simple scaling relations as a
function of $b$:
\begin{eqnarray}\label{ZFactPowers}
Z_\omega(b)&=&b^{z-2}\nonumber\\
Z_{\phi}(b)&=&b^{-2(z-2)+\eta}\nonumber\\
Z_{\tilde{\phi}}(b)&=&b^{-2(z-2)+\tilde{\eta}}.
\end{eqnarray}
These relations define three independent critical exponents of the
theory, $z$, $\eta$ and $\tilde{\eta}$. A further critical exponent
$\nu$ is associated with the positive eigenvalue of the linearized
RG equations around the fixed point.

\subsection{\label{Sse.ComReTime}Renormalization of the time and independent
critical exponents}

In dynamic RG procedures, there is in general a freedom to choose
some parameters constant while others are renormalized in a
non-trivial way. In the system described here, we choose to
renormalize the time (i.e. the frequency coordinate $\omega$ in
Fourier space) and to keep the parameters $c$ and $D$ invariant
under renormalization. In other cases, different choices are
commonly used. For example, renormalizing the dynamic model A with
$O(2)$-symmetry at thermodynamic equilibrium, the  coefficient $D$
is usually chosen to change under renormalization, while time is
simply rescaled \cite{baus76,taub92}. This model corresponds to the
particular case of the amplitude equation of the complex
Ginzburg-Landau theory where both $c_a$ and $u_a$ are equal to zero.
It is described by a dynamics of the form
\begin{equation}\label{modelA}
\partial_t\phi_{\alpha}=-D\frac{\delta\mathcal{H}}{\delta\phi_{\alpha}}+\zeta_{\alpha},
\end{equation}
and relaxes towards a thermodynamic equilibrium. Note, that since
this model satisfies an FD relation, the noise strength $D$ appears
as the mobility coefficient in the dynamics. Eq. (\ref{modelA})
shows that both choices, renormalizing the time or renormalizing
$D$, are equivalent. The complex Ginzburg-Landau theory discussed
here does not obey an FD relation. Therefore, a factorization of the
coefficient $D$ as it is done in Eq. (\ref{modelA}) would be
artificial in this case, and would enforce to redefine all other
parameters. Without this factorization, the structure of the theory
imposes to renormalize the time if the coefficient $c$ is kept
constant.

Note also that the absence of an FD relation in the theory changes
the structure of the RG equations as compared to an equilibrium
$O(2)$ model. Indeed, the FD relation imposes a constraint on the
renormalized quantities of the equilibrium theory, which can be
written as
\begin{equation}
Z_{\omega}\sqrt{Z_{\phi}Z_{\tilde{\phi}}^{-1}}\equiv 1,
\end{equation}
and which implies the following relation between the critical
exponents \cite{zinnB}:
\begin{equation}
z=2+\left(\tilde{\eta}-\eta\right)/2.
\end{equation}
In the non-equilibrium case considered here, such a constraint does
not exist and four truly independent critical exponents are present
in the theory as compared with three in the $O(2)$ dynamic model A.

\subsection{\label{Sse.1LoopRes}Results to one-loop order}

To one-loop order in perturbation theory, the $Z$ factors, as well
as the phase factor $\theta$ and the parameter $c_a$, are not
renormalized. We define the following reduced parameters
\begin{eqnarray}\label{RedVar}
&&\bar{r}=r/c\Lambda^2 \textrm{;}\quad \bar{\omega}_0=\omega_0/c\Lambda^2 \textrm{;} \quad \bar{c}_a=c_a/c\nonumber\\
&&\bar{u}=u/(c\Lambda^2)^2 \textrm{;} \quad \bar{u}_a=u_a/(c\Lambda^2)^2 \textrm{;}\nonumber\\
&&\mathcal{D}=\frac{4D\Lambda^d}{(4\pi)^{d/2}\Gamma(d/2)},
\end{eqnarray}
to express the RG flow in $d=4-\epsilon$. It is given by three
coupled equations
\begin{eqnarray}\label{RGEqWils}
\frac{d \bar{r}}{d l}&=&2\bar{r}+2\mathcal{D}\frac{\bar{u}}{1+\bar{r}}\nonumber\\
\frac{d \bar{u}}{d l}&=&\epsilon \bar{u}-\mathcal{D}\left[\frac{(\bar{u}^2-\bar{u}_a^2)(1+\bar{r})+2\bar{u}\bar{u}_a\bar{c}_a}{(1+\bar{r})
\left[\bar{c}_a^2+(1+\bar{r})^2\right]}+\frac{4\bar{u}^2}{(1+\bar{r})^2}\right]\nonumber\\
\frac{d \bar{u}_a}{d l}&=&\epsilon
\bar{u}_a+\mathcal{D}\frac{\bar{c}_a}{1+\bar{r}}\left[\frac{(\bar{u}^2-\bar{u}_a^2)(1+\bar{r})+2\bar{u}\bar{u}_a\bar{c}_a}{(1+\bar{r})
\left[\bar{c}_a^2+(1+\bar{r})^2\right]}\right]-6\mathcal{D}\frac{\bar{u}\bar{u}_a}{(1+\bar{r})^2},
\end{eqnarray}
and a fourth one associated with the renormalization of the
oscillation frequency:
\begin{equation}\label{Ome0EvolWils}
\frac{d \bar{\omega}_0}{d
l}=2\mathcal{D}\frac{\bar{u}_a}{1+\bar{r}}e^{-2l}.
\end{equation}
This last equation has to be integrated after the system
(\ref{RGEqWils}) has been solved.

\begin{figure}[h]
\scalebox{0.5}{
\includegraphics{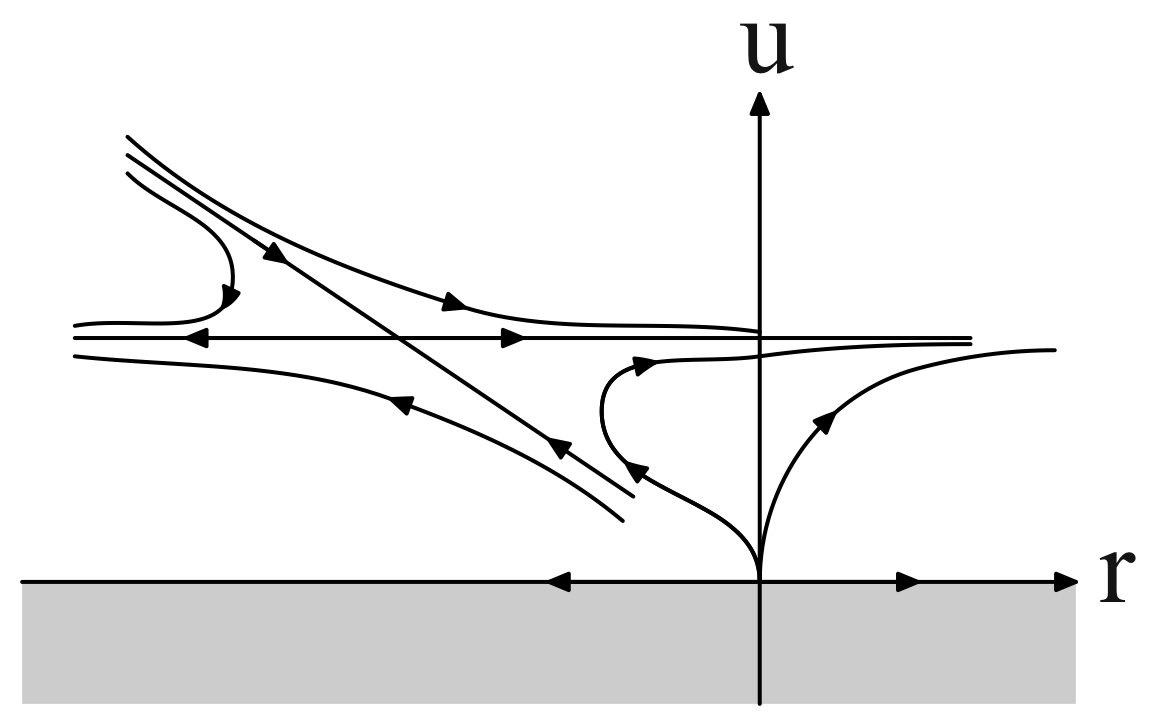}
}
\caption[fig1]{\label{Fig.FlowDiagWils}Qualitative representation of
RG flow to one-loop order in perturbation theory, projected on the
plane $(r,u)$ and for a space dimension $d<4$.}
\end{figure}

Since the coefficient $\bar{c}_a$ is not renormalized to first
order, we find one infrared stable fixed point for each of its
values. It is given by:
\begin{equation}\label{FPWils}
\bar{r}^*=-\frac{\epsilon}{5}, \quad
\bar{u}^*=\frac{\epsilon}{5\mathcal{D}} \quad \textrm{and} \quad
\bar{u}_a^*=\bar{c}_a\frac{\epsilon}{5\mathcal{D}}.
\end{equation}
Writing $\bar r=\bar r^*+\delta r$, $\bar u=\bar u^*+\delta u$ and
$\bar u_a=\bar u_a^*+\delta u_a$, the linearized RG equations at the
fixed point are given by:
\begin{equation}\label{LinRGWils}
\frac{d}{d l}
\left( \begin{array}{c}
\delta r \\
\delta u \\
\delta u_a \end{array}\right)
=
\left( \begin{array}{ccc}
2(1-\epsilon/5) & 2\mathcal{D} & 0 \\
0 & -\epsilon & 0 \\
0 & -4\bar{c}_a\epsilon/5 & -\epsilon/5 \end{array}\right)
\left( \begin{array}{c}
\delta r \\
\delta u \\
\delta u_a \end{array}\right).
\end{equation}
The RG flow of the theory is three dimensional. We show in Fig.
\ref{Fig.FlowDiagWils} the qualitative RG flow projected on the
plane $(r,u)$.
Finally the critical exponents to the one-loop order
read:
\begin{equation}
\nu=\frac{1}{2}+\frac{\epsilon}{10} ; \quad z=2 ; \quad \eta=1 ;
\quad \tilde{\eta}=0.
\end{equation}

To one-loop order, the three $Z$ factors and the parameter $c_a$ are
not renormalized. Therefore, these calculations are insufficient to
determine the fixed point value of $c_a$ as well as three of the
four independent critical exponents of the theory. Calculations to
two-loop order are necessary to obtain the full RG structure and the
critical properties. Wilson's momentum shell RG scheme is not well
suited to perform such calculations and it is technically far more
convenient to perform these using a Callan-Symanzik RG scheme,
adapted to the renormalization of critical oscillators. This is
discussed in the next section.

\section{\label{Se.RG} Callan-Symanzik RG scheme}

The Callan-Symanzik RG scheme avoids the introduction of a cut-off
in the momentum space, which is responsible for making the
evaluation of multiple integrals technically difficult in Wilson's
scheme. In its absence, the calculation of two-loop and higher order
Feynman integrals is easier. Here we present the general structure
of the Callan-Symanzik RG scheme adapted to the study of coupled
oscillators. We discuss the RG flow to two-loop order in
perturbation theory and show that the Callan-Symanzik RG scheme
described here is consistent with the momentum shell procedure
presented before.

\subsection{\label{Sse.GenRGForm} General formalism}

Within a Callan-Symanzik RG scheme \cite{brez76c,zinnB}, we start
from a bare theory that follows a dynamics of the form
(\ref{MatEvol}), whose parameters are labeled with a superscript
``0''. We then define the renormalized theory such that its
effective action is of the form (\ref{ActPhi}). This requires to
introduce a phase shift $\delta\theta$ and a frequency shift
$\delta\omega_0$ between the bare fields
$(\phi_{\alpha}^0,\tilde{\phi}_{\alpha}^0)$ and the renormalized
fields $(\phi_{\alpha},\tilde{\phi}_{\alpha})$, such that
\begin{eqnarray}\label{d.Phi/d.Omega}
\phi^0_{\alpha}(\mathbf{x},t^0)&=&\Omega_{\alpha\beta}(-\delta\omega_0 t)\,Z_{\phi}^{1/2}Z_{\omega}\,\phi_{\beta}(\mathbf{x},t)\nonumber\\
\tilde{\phi}^0_{\alpha}(\mathbf{x},t^0)&=&\Omega_{\alpha\beta}(-\delta\theta-\delta\omega_0
t)\,Z_{\tilde{\phi}}^{1/2}Z_{\omega}\,\tilde{\phi}_{\beta}(\mathbf{x},t).
\end{eqnarray}
Here we have introduced $Z$ factors for the renormalization of the
fields and the time ($t^0=Z_{\omega}^{-1}t$). We furthermore
introduce dimensionless coupling constants $g$ and $g_a$ and a scale
parameter $\mu$ with $u=\mu^{\epsilon}(4\pi)^{-\epsilon/2}g$ and
$u_a=\mu^{\epsilon}(4\pi)^{-\epsilon/2}g_a$. Depending on $\mu$, we
relate the bare quantities  to the renormalized ones by additional
$Z$ factors: $g^0=Z_g(Z_{\tilde{\phi}}Z_{\phi}^3)^{-1/2}g$,
$g^0_a=Z_{g_a}(Z_{\tilde{\phi}}Z_{\phi}^3)^{-1/2}g_a$,
$r^0=r_c^0+Z_r r$ and $ c^0_a=Z_{c_a}c_a$. The dependence of the
renormalized parameters $g$, $g_a$ and $c_a$ on $\mu$ defines three
beta functions. Denoting $\vec{g}=\left(g,g_a,c_a\right)$, we write
$\vec{\beta}\left(\vec{g},\epsilon\right)=\mu(\partial_{\mu}
\vec{g})_0$, where
$\vec{\beta}=\left(\beta,\beta_a,\beta_{c}\right)$ and
$(\partial_{\mu})_0$ denotes differentiation with fixed $u^0$,
$u^0_a$ and $c^0_a$. Note again that $c$ and $D$ are kept constant
under renormalization and that we choose units such that $c=1$ in
the following. The renormalized correlation and response functions
$G_{\alpha\beta}=\langle\phi_\alpha\phi_\beta\rangle_c$ and
$\gamma_{\alpha\beta}=\langle\phi_\alpha\tilde{\phi}_\beta\rangle_c$
are related to the physical observables $C_{\alpha\beta}$ and
$\chi_{\alpha\beta}$ via
\begin{eqnarray}\label{Phys/RenCoRespFct}
C_{\alpha\beta}(\mathbf{x},t^0)&=&\Omega_{\alpha\sigma}\left(-\omega_0\,t\right)
Z_{\phi}Z_{\omega}^{2}\,G_{\sigma\beta}(\mathbf{x},t)\nonumber\\
\chi_{\alpha\beta}(\mathbf{x},t^0)&=&\Lambda^{-1}\,\Omega_{\alpha\sigma}\left(\theta-\omega_0\,t\right)
(Z_{\phi}Z_{\tilde{\phi}})^{1/2}Z_{\omega}^{2}\,\gamma_{\sigma\beta}(\mathbf{x},t).
\end{eqnarray}
The frequency $\omega_0=Z_{\omega}^{-1}\omega_0^0+\delta\omega_0$
and the phase $\theta=\theta^0+\delta\theta$ are renormalized
according to
\begin{eqnarray}\label{EvolOmeThetaEffCS}
\mu(\partial_{\mu}\delta\theta)_0&=&\gamma_{\theta}(g,g_a,c_a,\epsilon)\nonumber\\
\mu(\partial_{\mu}\delta\omega_0)_0&=&r\,\gamma_{\omega_0}(g,g_a,c_a,\epsilon),
\end{eqnarray}
which defines the Wilson's functions $\gamma_\theta$ and
$\gamma_{\omega_0}$. In addition, we define the Wilson's functions
associated with the dependence of the $Z$ factors on $\mu$:
$\gamma_r=\mu(\partial_{\mu}\ln{Z_r})_0$,
$\gamma_{\omega}=\mu(\partial_{\mu}\ln{Z_{\omega}})_0$,
$\gamma=\mu(\partial_{\mu}\ln{Z_{\phi}})_0$ and
$\tilde{\gamma}=\mu(\partial_{\mu}\ln{Z_{\tilde{\phi}}})_0$.

The independence of the bare theory with respect to the scale
parameter $\mu$ leads to the Callan-Symanzik equations, which we
write in terms of the renormalized theory:
\begin{equation}\label{CSEq}
\Bigg[
\frac{\partial}{\partial\ln{\mu}}+\vec{\beta}\cdot\frac{\partial}{\partial
\vec{g}}-\sum_{i(j)=1}^{\tilde{N}(N)}\gamma_{\omega}\,
\frac{\partial}{\partial\ln
\omega_{i(j)}}-\gamma_r\frac{\partial}{\partial\ln{r}}\Bigg]\Gamma^{(\tilde{N},N)}
=\Bigg[(\tilde{N}+N-1)\gamma_{\omega}+\frac{\tilde{N}}{2}\tilde{\gamma}+\frac{N}{2}\gamma\Bigg]\Gamma^{(\tilde{N},N)}.
\end{equation}
Here $\Gamma^{(\tilde{N},N)}$ is the vertex function with
$\tilde{N}$ and $N$ truncated external legs corresponding to the
fields $\tilde{\phi}$ and $\phi$, respectively. It is a function of
$\tilde N$ variables $(\omega_i,{\bf q_i})$ and $N$ variables
$(\omega_j,{\bf q_j})$ describing the frequencies and wavelengths
associated with all external legs, and depends on the renormalized
set of parameters.

In order to calculate the Wilson's beta and gamma functions which
appear in the Callan-Symanzik equations (\ref{CSEq}), we decompose
the bare action associated with the fields $\phi_{\alpha}^0$ and
$\tilde{\phi}_{\alpha}^0$ as
\begin{equation}\label{ActDecCS}
\mathcal{S}^0\left[\tilde{\phi}^0_{\alpha},\phi^0_{\alpha}\right]=\mathcal{S}_{\rm{R}}\left[\tilde{\phi}_{\alpha},\phi_{\alpha}\right]
+\delta\mathcal{S}\left[\tilde{\phi}_{\alpha},\phi_{\alpha}\right]+\mathcal{S}_{\rm{Mop}}\left[\tilde{\phi}_{\alpha},\phi_{\alpha}\right].
\end{equation}
Here $\mathcal{S}_{\rm{R}}[\tilde{\phi}_{\alpha},\phi_{\alpha}]$
represents the action of the renormalized theory,
$\mathcal{S}_{\rm{Mop}}[\tilde{\phi}_{\alpha},\phi_{\alpha}]$ the
action associated with the ``mass operator''
$[\tilde{\phi}_{\alpha}\phi_{\beta}]$, and
$\delta\mathcal{S}[\tilde{\phi}_{\alpha},\phi_{\alpha}]$ combines
the counter-terms. The integrals corresponding to the Feynman
diagrams of the theory contain poles as a function of the small
dimensional parameter $\epsilon$. The $Z$ factors are determined
such that the counter-terms absorb these poles and the effective
action is finite. We write:
\begin{eqnarray}\label{DiffActCS1}
\mathcal{S}_{\rm{R}}\left[\tilde{\phi}_{\alpha},\phi_{\alpha}\right]&=&\int_{\mathbf{q},\omega}\left\{
D\tilde{\phi}_{\alpha}\tilde{\phi}_{\alpha}-\tilde{\phi}_{\alpha}\left[i\omega+c\mathbf{q}^2\right]\phi_{\alpha}
-\tilde{\phi}_{\alpha}\left[c_a\mathbf{q}^2\right]\varepsilon_{\alpha\beta}\phi_{\beta}\right\}\nonumber\\
&&-\int_{k_1k_2k_3}\mu^{\epsilon}(4\pi)^{-\epsilon/2}g_{\alpha\beta}\tilde{\phi}_{\alpha}\phi_{\beta}\phi_{\gamma}\phi_{\gamma}
\end{eqnarray}
\begin{eqnarray}\label{DiffActCS2}
\delta\mathcal{S}\left[\tilde{\phi}_{\alpha},\phi_{\alpha}\right]&=&\int_{\mathbf{q},\omega}\left\{
D\left(Z_{\omega}Z_{\tilde{\phi}}-1\right)\tilde{\phi}_{\alpha}\tilde{\phi}_{\alpha}-\tilde{\phi}_{\alpha}\left[
i\omega\left(Z_{\omega}^2\sqrt{Z_{\phi}Z_{\tilde{\phi}}}\cos{\delta\theta}-1\right)\right.\right.\nonumber\\
&&\left.\left.+c\mathbf{q}^2\left(Z_{\omega}\sqrt{Z_{\phi}Z_{\tilde{\phi}}}\cos{\delta\theta}
-Z_{\omega}\sqrt{Z_{\phi}Z_{\tilde{\phi}}}\,Z_{c_a}c_a/c\,\sin{\delta\theta}-1\right)\right]\phi_{\alpha}\right.\nonumber\\
&&\left.-\tilde{\phi}_{\alpha}\left[i\omega\left(Z_{\omega}^2\sqrt{Z_{\phi}Z_{\tilde{\phi}}}\sin{\delta\theta}\right)\right.\right.\nonumber\\
&&\left.\left.+c_a\mathbf{q}^2\left(Z_{\omega}\sqrt{Z_{\phi}Z_{\tilde{\phi}}}Z_{c_a}\cos{\delta\theta}
+Z_{\omega}\sqrt{Z_{\phi}Z_{\tilde{\phi}}}\,c/c_a\,\sin{\delta\theta}-1\right)\right]\varepsilon_{\alpha\beta}\phi_{\beta}\right\}\nonumber\\
&&-\mu^{\epsilon}(4\pi)^{-\epsilon/2}\int_{k_1k_2k_3}\left\{\left[(Z_gZ_{\omega}^3\cos{\delta\theta}-1)g
-Z_{g_a}Z_{\omega}^3\sin{\delta\theta}\,g_a\right]\tilde{\phi}_{\alpha}\phi_{\alpha}\phi_{\gamma}\phi_{\gamma}\right.\nonumber\\
&&\hspace{3cm}
\left.+\left[(Z_{g_a}Z_{\omega}^3\cos{\delta\theta}-1)g_a+Z_gZ_{\omega}^3\sin{\delta\theta}\,g\right]
\varepsilon_{\alpha\beta}\tilde{\phi}_{\alpha}\phi_{\beta}\phi_{\gamma}\psi_{\gamma}\right\}\nonumber\\
\end{eqnarray}
\begin{eqnarray}\label{DiffActCS3}
\mathcal{S}_{\rm{Mop}}\left[\tilde{\phi}_{\alpha},\phi_{\alpha}\right]&=&\int_{\mathbf{q},\omega}\left\{
-\tilde{\phi}_{\alpha}\phi_{\alpha}\sqrt{Z_{\phi}Z_{\tilde{\phi}}}Z_{\omega}(rZ_r\cos{\delta\theta})\right\}\nonumber\\
&&+\int_{\mathbf{q},\omega}\left\{
-\tilde{\phi}_{\alpha}\varepsilon_{\alpha\beta}\phi_{\beta}\sqrt{Z_{\phi}Z_{\tilde{\phi}}}Z_{\omega}
\left[-Z_{\omega}\delta\omega_0+rZ_r\sin{\delta\theta}\right]\right\},
\end{eqnarray}
where
\begin{equation}\label{CouplConstGMat}
g_{\alpha\beta}=\left( \begin{array}{cc}
g & -g_{a}\\
g_{a} & g
\end{array} \right).
\end{equation}

Having determined the $Z$ factors, we can calculate the beta
functions. Writing
\begin{eqnarray}
\beta(g,g_a,c_a,\epsilon)&=&-\epsilon g+\beta^{(4)}(g,g_a,c_a)\nonumber\\
\beta_a(g,g_a,c_a,\epsilon)&=&-\epsilon g_a+\beta_a^{(4)}(g,g_a,c_a)\nonumber\\
\beta_c(g,g_a,c_a,\epsilon)&=&\beta_c^{(4)}(g,g_a,c_a),
\end{eqnarray}
the functions $\beta^{(4)}$ do not depend on $\epsilon$ and are
given by
\begin{equation}\label{BetaFctExpr}
\left( \begin{array}{c}
\beta ^{(4)}\\
\beta_a^{(4)} \\
\beta_c^{(4)}
\end{array}\right)
=
\left( \begin{array}{ccc}
g\frac{\partial\bar{Z}_g}{\partial g} & g\frac{\partial\bar{Z}_g}{\partial g_a} & g\frac{\partial\bar{Z}_g}{\partial c_a}\\
g_a\frac{\partial\bar{Z}_{g_a}}{\partial g} &
g_a\frac{\partial\bar{Z}_{g_a}}{\partial g_a} & g_a\frac{\partial\bar{Z}_{g_a}}{\partial c_a} \\
c_a\frac{\partial {Z}_{c_a}}{\partial g} & c_a\frac{\partial
{Z}_{c_a}}{\partial g_a} & {Z}_{c_a}+c_a\frac{\partial
{Z}_{c_a}}{\partial c_a}
\end{array}\right)^{(1)}
\left( \begin{array}{c}
g \\
g_a \\
0 \end{array}\right),
\end{equation}
where the superscript ``$(1)$'' indicates the coefficients of the
poles in $\epsilon^{-1}$ of the different $Z$ factors in the matrix.
Here $\bar{Z}_g=Z_g(Z_{\tilde{\phi}}Z_{\phi}^3)^{-1/2}$ and
$\bar{Z}_{g_a}=Z_{g_a}(Z_{\tilde{\phi}}Z_{\phi}^3)^{-1/2}$. The
Wilson's gamma functions of the theory are independent of
$\epsilon$. We have
\begin{equation}\label{GamFctExpr}
\gamma_r(g,g_a,c_a)=-g\frac{\partial Z_r^{(1)}}{\partial
g}-g_a\frac{\partial Z_r^{(1)}}{\partial g_a},
\end{equation}
and corresponding expressions for $\gamma_{\omega}$, $\gamma$ and
$\tilde{\gamma}$. Furthermore,
\begin{eqnarray}\label{GamFctHopfExpr}
\gamma_{\theta}(g,g_a,c_a)&=&-g\frac{\partial
\delta\theta^{(1)}}{\partial g}-g_a\frac{\partial
\delta\theta^{(1)}}{\partial g_a}\nonumber\\
\gamma_{\omega_0}(g,g_a,c_a)&=&\frac{1}{r}\left(-g\frac{\partial
\delta\omega_0^{(1)}}{\partial g}-g_a\frac{\partial
\delta\omega_0^{(1)}}{\partial g_a}\right).
\end{eqnarray}

The RG fixed points correspond to the values $\vec{g}^*$ of
$\vec{g}$ for which the beta functions are zero. The critical
exponents of the theory are given by the fixed point values of the
gamma functions. We have:
\begin{eqnarray}\label{CritExpCS}
\nu=\big(2+\gamma_r(\vec{g}^*)\big)^{-1}&&
z=2+\gamma_{\omega}(\vec{g}^*)\nonumber\\
\eta=\gamma(\vec{g}^*)+2\gamma_{\omega}(\vec{g}^*)&&
\tilde{\eta}=\tilde{\gamma}(\vec{g}^*)+2\gamma_{\omega}(\vec{g}^*).
\end{eqnarray}

\subsection{\label{Sse.OneLoop}One-loop order calculations}

As in the framework of Wilson's momentum shell integration scheme,
to one-loop order in perturbation theory, only the parameters
$\omega_0$, $r$, $g$ and $g_a$ are renormalized. We find
$\gamma_r=-2\mathcal{D}g$ and $\gamma_{\omega_0}=2\mathcal{D}g_a$,
where $\mathcal{D}=4D/(4\pi)^2$ (see Appendix \ref{Ap.ExplExpr} for
details about the calculations). Furthermore, the beta functions
associated with the renormalization of $g$ and $g_a$ read
\begin{eqnarray}\label{BandBaFctCS}
\beta
&\simeq&-\epsilon g-\mathcal{D}\left[\frac{g_a^2-g^2-2gg_a {c}_a}{1+{c}_a^2}-4g^2\right]\nonumber\\
\beta_a &\simeq&-\epsilon
g_a-\mathcal{D}\left[\frac{{c}_a}{1+{c}_a^2}(g^2-g_a^2+2gg_a
{c}_a)-6gg_a\right].
\end{eqnarray}
We find the same results as those discussed in Section V. Choosing
$c_a$ as a parameter, the fixed points read:
\begin{equation}\label{FPgga(ca)CS}
g^*\simeq \,\epsilon/5\mathcal{D}, \quad g_a^*\simeq
\,c_a\,\epsilon/5\mathcal{D}.
\end{equation}

\subsection{\label{Sse.TwoLoop}Two-loop order calculations and universality
class}

\begin{figure}[h]
\scalebox{0.4}{
\includegraphics{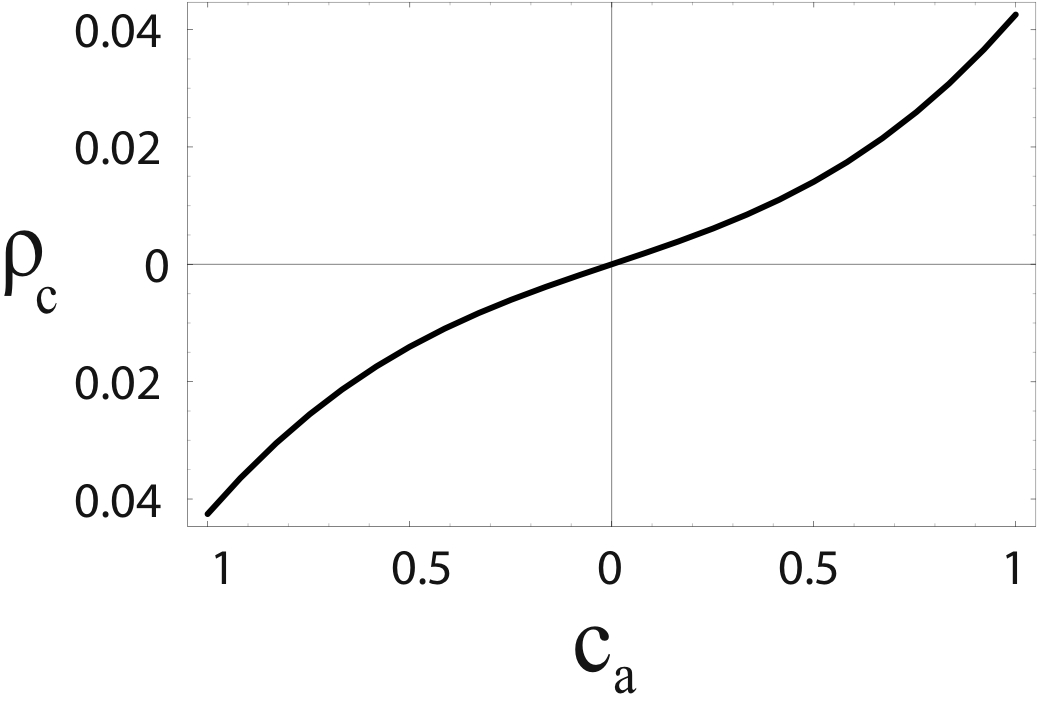}
}

\caption[fig2]{\label{Fig.RhoCa} The renormalization of the
parameter $c_a$ is described by the Wilson's function $\rho_c$,
whose expression to two-loop order is given by Eq. (\ref{rho_c}) and
which is displayed here as a function of $c_a$. A fixed point of the
theory is characterized by $\rho_c=0$. A single fixed point exists
for $c_a^*=0$.}
\end{figure}

To this order in perturbation theory, all Z factors and parameters
of the theory are renormalized. Details about this renormalization
procedure are given in the appendices: Appendix A gives graphic
representations of the Feynman diagrams of the perturbation theory,
Appendix B gives the main calculations steps and explicit
expressions for the associated integrals, and Appendix C provides
expressions for the different Wilson's functions of the theory,
including $\beta_c(g,g_a,c_a)$. The renormalization of the
propagator to second order does not affect the first-order results
discussed above. The fixed point condition for the full problem can
thus be written in the form $\rho_c(c_a^*)=0$, where
\begin{equation}\label{Rho_c}
\rho_c(c_a)=\epsilon^{-2}\beta_c(g^*,g_a^*,c_a),
\end{equation}
and $c_a^*$ denotes the fixed point value of $c_a$. Here, we used
the fixed point values $g^*$ and $g_a^*$ given by Eq.
(\ref{FPgga(ca)CS}). Note that to two-loop order in perturbation
theory, this function $\rho_c(c_a)$ is independent of $\epsilon$.
The full expression of the function $\rho_c(c_a)$ is given in
appendix C Eq. (\ref{rho_c}), and its graphic representation in Fig.
\ref{Fig.RhoCa}.

A single fixed point of the theory exists with $c_a^*=0$ and
$g_a^*=0$. This fixed point is IR-stable. It is the same as the one
of the real Ginzburg-Landau theory with $O(2)$ symmetry. As a
consequence, the dominant critical exponents are the same as those
known for the $O(2)$ dynamic model A. They are
$\nu\simeq1/2+\epsilon/10$, $\eta\simeq\epsilon^2/50$ and
$z\simeq2+\epsilon^2\left(6\ln(4/3)-1\right)/50$, associated with
the relation $z=2+(\tilde{\eta}-\eta)/2$. Furthermore, at this fixed
point, both functions $\gamma_{\theta}$ and $\gamma_{\omega_0}$ are
equal to zero. Therefore, at the critical point, the effective phase
and frequency become scale-invariant. Note however that since the
renormalized fields differ from the physical fields by
time-dependent variable transformations, the physical correlation
and response functions differ significantly from those associated
with the $O(2)$ dynamic model. The asymptotic expressions of these
functions are discussed is Section \ref{Se.AsymCritBehav}.

The critical behaviors of the theory are characterized by the
linearized RG flow in the vicinity of the fixed point. The
linearized flow equations can be written as
\begin{equation}
\frac{d \vec{g}}{d
\ln{\mu}}=\underline{\underline{\omega}}\cdot\vec{g}(\mu),
\end{equation}
where we have introduced the matrix
\begin{equation}\label{Eq.OmeMat}
\underline{\underline{\omega}}=
\left( \begin{array}{ccc}
\epsilon & 0 & 0 \\
0 & \epsilon/5 & -\epsilon^2/25\mathcal{D} \\
0 & 0 & \epsilon^2/50 \end{array}\right).
\end{equation}
The eigenvalues of this matrix are:
\begin{equation}\label{Eq.OmeMatEV}
\omega_1=\varepsilon \quad \textrm{;} \quad
\omega_2=\frac{\varepsilon}{5} \quad \textrm{;} \quad
\omega_3=\frac{\varepsilon^2}{50}.
\end{equation}
In addition to $\omega_1$, which is known from the $O(2)$ symmetric
dynamic model A, we find here two new universal quantities
$\omega_2$ and $\omega_3$, which are specific to critical
oscillators.

\subsection{\label{Sse.FlowDiag}Flow diagram of the theory to two-loop order
in perturbation theory}

\begin{figure}[h]
\scalebox{0.4}{
\includegraphics{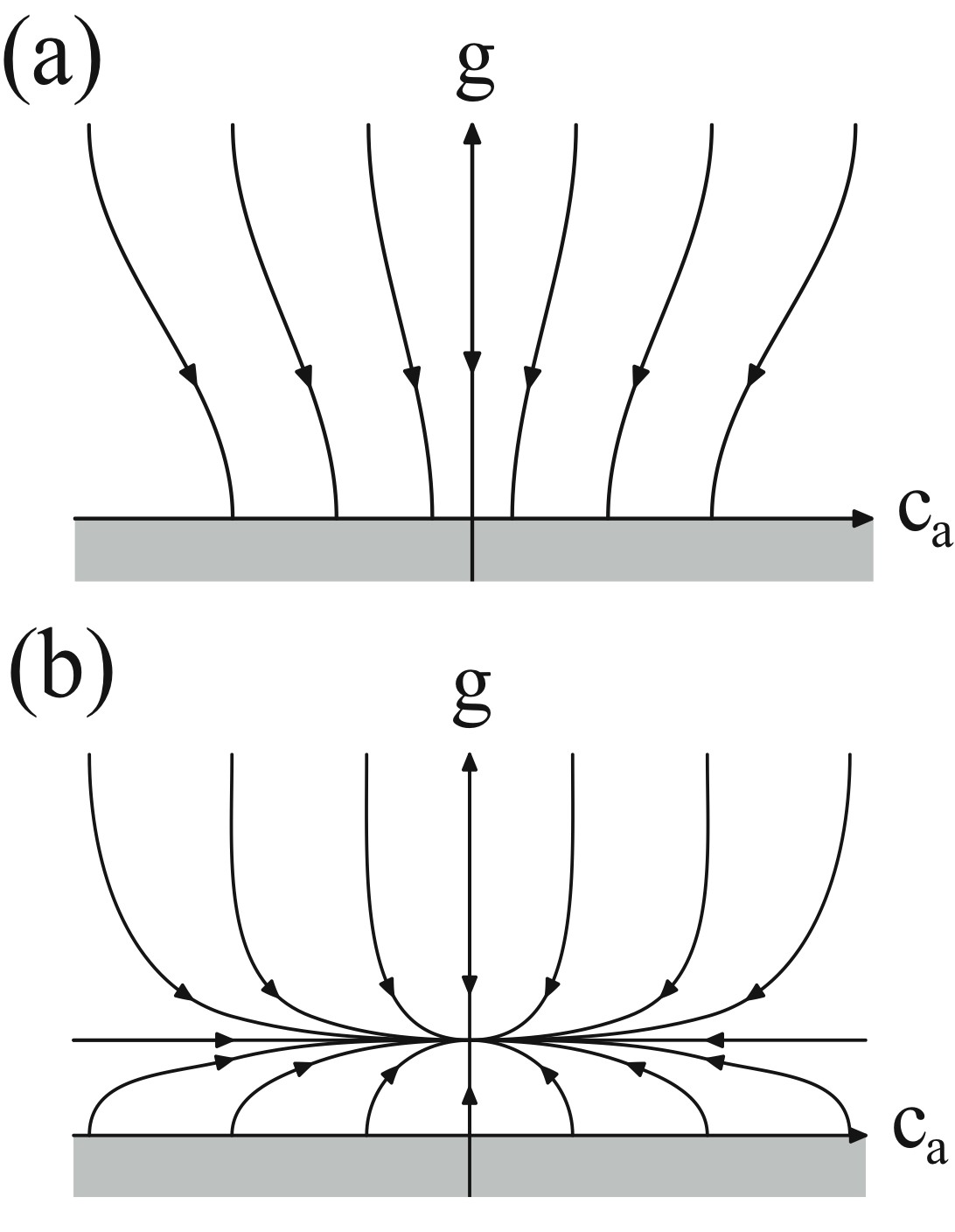}
}

\caption[fig3]{\label{Fig.FlowDiagCS} Schematic representation of
the RG flow of the theory, obtained to two-loop order in
perturbation theory, and projected on the plane $(g,c_a)$. (a) Space
dimension $d>4$. (b) Space dimension $d<4$.}
\end{figure}

The RG flow of the theory within the Callan-Symanzik RG scheme is
given by the variations of the three parameters $g$, $g_a$ and $c_a$
under renormalization. We display in Fig. \ref{Fig.FlowDiagCS} the
projection of this flow on the plane $(g,c_a)$ for two different
space dimensions, above and below the upper critical dimension
$d_c=4$.
The first plot corresponds to $d>4$ and the second to
$d<4$.

Because the RG flow here is defined in an enlarged space, its
structure differs remarkably from the one of the real
Ginzburg-Landau theory. For $d>4$, we find a line of Gaussian fixed
points corresponding to $g^*=0$ and $g_a^*=0$ for any value of
$c_a^*$. These fixed points characterize the mean field universality
classes of critical oscillators, which depend on the value of
$c_a^*$. Below the critical dimension $d_c=4$, a single fixed point
exists with $g^*=\epsilon/5\mathcal{D}$, $g_a^*=0$ and $c_a^*=0$.
Because of the existence of a whole line of Gaussian fixed points
that change their stability at $d=d_c$, the RG flow has a singular
structure near $d=d_c$. As a consequence, the large scale behaviors
of critical oscillators for $d>d_c$ can vary correspondingly to
different values of the effective parameter $c_a$. For $d<d_c$
however, the characteristic critical behaviors are always described
by the single fixed point with $c_a^*=0$, relevant for this case.

\section{\label{Se.AsymCritBehav}  Correlation and response functions
and violation of the fluctuation-dissipation relation}

In the previous section, we have discussed the RG flow and fixed
point structures of the theory. We have seen that the renormalized
fields are described in a reference frame that oscillates with
renormalized frequency and phase factors. The fixed point theory is
formally equivalent to the one of a critical point at thermodynamic
equilibrium, namely the critical point of the dynamic
Ginzburg-Landau theory with an $O(2)$ symmetry. However, we show now
that the correlation and response functions of the physical fields
studied here have different properties.

\subsection{\label{Sse.CorrResp} Asymptotic behaviors of the correlation and response functions
in the critical regime}

The asymptotic behaviors of the correlation and response functions
of the theory near criticality can be determined using the RG flow
and applying a matching procedure to link these functions with their
expressions off criticality (see e.g. \cite{rudn76}). In the present
case, such a matching procedure needs to be adapted. Indeed, the
physical correlation and response functions are related to those
defined for the renormalized fields by the time-dependent
transformations (\ref{Phys/RenCoRespFct}). We therefore have to add
to the usual matching procedure a scale-dependent transformation to
describe the physical theory in its original reference frame. This
transformation depends on the effective frequency and phase of the
oscillators, which are renormalized by the RG procedure. Taking all
this into account, we can write effective asymptotic expressions for
the functions associated with physical quantities.

The effective linear response function of the physical theory
behaves as (for $q\xi\gg 1$ and for stimulation at the effective
frequency $\omega_0^{\rm eff}$)
\begin{equation}\label{EffRespFct_q}
\chi(q,\omega=\omega_0^{\rm{eff}})\simeq\frac{1}{q^{2-\eta}}
\,\frac{1}{2\Lambda_{\rm{eff}}}\left[\frac{e^{i\theta(q)}}{c_{\rm{eff}}+i\gamma(q)}\right].
\end{equation}
Here, we denote by $\omega_0^{\rm eff}$ the effective oscillation
frequency at the bifurcation, and by $\xi$ the correlation length in
the non-oscillating phase. Furthermore, we have introduced the
functions $\theta(q)\simeq \theta_{\rm{eff}}+
\alpha_{\rm{eff}}q^{\omega_2}+\beta_{\rm{eff}}q^{\omega_3}$ and
$\gamma(q)\simeq \gamma_{\rm{eff}}q^{\omega_3}$ of the wave number
$q=|\mathbf{q}|$, as well as non-universal effective quantities
denoted by the index ``eff''. These functions are derived
respectively from the renormalizations of the parameters $\theta$
and $c_a$ in the vicinity of the fixed point. Note that they depend
on the universal critical exponents given by Eq.
(\ref{Eq.OmeMatEV}). Similarly to Eq. (\ref{EffRespFct_q}), the
correlation function behaves as
\begin{equation}
C(q,\omega=\omega_0^{\rm{eff}})\simeq\frac{1}{q^{z+2-\eta}}\,\frac{D_{\rm{eff}}}{c_{\rm{eff}}^2+\gamma(q)^2}.
\end{equation}
Related expressions can be obtained for the frequency dependence for $q=0$
in the regime $(\omega-\omega_0^{\rm{eff}})\xi^z\gg 1$.
They are given by
\begin{eqnarray}\label{EffCorrRespOm}
\chi(q=0,\omega)&\simeq&\pm i\,\frac{e^{i\theta(\omega-\omega_0^{\rm{eff}})}}{2\Lambda_{\rm{eff}}}\,
\frac{1}{\left|\omega-\omega_0^{\rm{eff}}\right|^{\frac{2-\eta}{z}}[.]^{\frac{2-\eta}{z}}}\nonumber\\
C(q=0,\omega)&\simeq&\frac{D_{\rm{eff}}}{\left|\omega-\omega_0^{\rm{eff}}\right|^{\frac{2+z-\eta}{z}}[.]^{\frac{2+z-\eta}{z}}},
\end{eqnarray}
where
\begin{eqnarray}\label{ReOm0}
&&[.] \simeq
\left[1+\rho_{\rm{eff}}\left|\omega-\omega_0^{\rm{eff}}\right|^{\frac{\omega_2}{2}}
+\sigma_{\rm{eff}}\left|\omega-\omega_0^{\rm{eff}}\right|^{\frac{1}{2\nu}-1}\right]\nonumber\\
&&\theta(\omega-\omega_0^{\rm{eff}}) \simeq \theta_{\rm{eff}}+
\alpha_{\rm{eff}}\left|\omega-\omega_0^{\rm{eff}}\right|^{\frac{\omega_2}{z}}
+\beta_{\rm{eff}}\left|\omega-\omega_0^{\rm{eff}}\right|^{\frac{\omega_3}{z}},
\end{eqnarray}
and where ``$\pm$'' corresponds to $\omega-\omega_0^{\rm{eff}}$
being positive or negative, respectively. The anomalous dependences
on frequencies as given by Eq. (\ref{ReOm0}) are due to the
non-trivial evolutions of the parameters $\theta$ and $\omega_0$
under renormalization.

\subsection{\label{Sse.GenFDT} Generalized fluctuation-dissipation relation in the critical regime}

A collective system close to a Hopf bifurcation operates far from
thermodynamic equilibrium. Therefore, the correlation function
$C_{\alpha\beta}$ and the linear response function
$\chi_{\alpha\beta}$ do not obey the fluctuation-dissipation (FD)
relation that is characteristic of thermodynamic equilibrium.
Interestingly, the effective theory at the RG fixed point, expressed
in terms of the renormalized fields $\phi_\alpha$  and $\tilde
\phi_\alpha$, is formally equivalent to a fixed point theory at
thermodynamic equilibrium. Therefore, exactly at the fixed point, a
relation appears between the correlation and response functions
$G_{\alpha\beta}$ and $\gamma_{\alpha\beta}$ of the renormalized
fields. It takes the form
\begin{equation}
G_{\alpha\beta}=\frac{2D}{\omega}\gamma''_{\alpha\beta},
\end{equation}
where
$\gamma_{\alpha\beta}=\gamma'_{\alpha\beta}+i\gamma''_{\alpha\beta}$
has been split into its real and imaginary parts. The emergence of
this relation in the critical regime can be discussed by the
introduction of the function \footnote{Note that $G_{11}>0$ so that
$F$ can always be defined.}
\begin{equation}\label{Eq.FDRel}
F(\mathbf{q},\omega)=\frac{2D}{\omega}\frac{
\gamma''_{11}(\mathbf{q},\omega)}{G_{11}(\mathbf{q},\omega)}.
\end{equation}
The evolution of this quantity under renormalization is described by
the following Callan-Symanzik equation:
\begin{equation}
\left[\frac{\partial}{\partial\ln{\mu}}+\vec{\beta}\cdot\frac{\partial}{\partial
\vec{g}}-\gamma_{\omega}\frac{\partial}{\partial\ln{\omega}}-\gamma_r\frac{\partial}{\partial\ln{r}}\right]F
=\left[\gamma_{\omega}+\frac{1}{2}(\gamma-\tilde{\gamma})\right]F.
\end{equation}
Since the fixed point theory obeys the FD relation, we have
$F(\mathbf{q},\omega,r,\vec{g}^*,\mu)=1$. Note that in mean field
theory we find $F^{\rm{mf}}=(\omega^2+R^2-(c_a
q^2)^2)/(\omega^2+R^2+(c_a q^2)^2)$, which differs from $F=1$ if
$c_a\neq 0$.

The fact that the renormalized theory at the fixed point obeys the
FD relation $F=1$, implies that the correlation and response
functions $C_{\alpha\beta}$ and $\chi_{\alpha\beta}$ at that point
are not independent. Since they are related to $G_{\alpha\beta}$ and
$\gamma_{\alpha\beta}$ by Eq. (\ref{Phys/RenCoRespFct}), we find
\begin{eqnarray}\label{nFD}
\cos{\theta_{\rm{eff}}}\chi''_{11}+\sin{\theta_{\rm{eff}}}\chi''_{12}&=&\frac{1}{2\Lambda_{\rm{eff}}D_{\rm{eff}}}\left(\omega
C_{11}+i\omega_0^{\rm{eff}}C_{12}\right)\nonumber\\
\cos{\theta_{\rm{eff}}}\chi'_{12}-\sin{\theta_{\rm{eff}}}\chi'_{11}&=&\frac{1}{2\Lambda_{\rm{eff}}D_{\rm{eff}}}\left(\omega_0^{\rm{eff}}
C_{11}+i\omega C_{12}\right).
\end{eqnarray}
Here again
$\chi_{\alpha\beta}=\chi_{\alpha\beta}'+i\chi_{\alpha\beta}''$ has
been separated in its real and imaginary parts. At the bifurcation,
this relation is asymptotically satisfied in the long time and
wave-length limits. It is a consequence of symmetry properties of
the fixed point theory which impose constraints on the correlation
and response functions at criticality. Indeed, the FD relation is
connected with time-reversal invariance, which emerges for the
fields $\phi_\alpha$ and $\tilde \phi_\alpha$ at criticality while
it is not obeyed for the physical fields $\psi_\alpha$ and $\tilde
\psi_\alpha$.

\subsection{\label{Sse.UnivBreakFD}  Breaking of the fluctuation-dissipation relation}

The relation between the physical correlation and response functions
at criticality given by Eq. (\ref{nFD}) is not an FD relation. In
order to characterize the violation of the FD relation between
$C_{\alpha\beta}$ and $\chi_{\alpha\beta}$, we define an effective
temperature $T_{\rm eff}$, which depends on frequency and wave
vector \cite{mart01b}:
\begin{equation}
\frac{T_{\rm eff}(\mathbf{q},\omega)}{T}=\frac{\omega}{2 k_B
T}\,\frac{C_{11}(\mathbf{q},\omega)}{\chi''_{11}(\mathbf{q},\omega)}.
\end{equation}
Here, $k_B$ denotes the Boltzmann constant and $T$ is the
temperature. Using the previous asymptotic expressions for the
two-point correlation and response functions, we find universal
behaviors of this effective temperature at criticality:
\begin{eqnarray}\label{EffTempDiv}
T_{\rm eff}(q,\omega=\omega_0^{\rm{eff}})/T&\sim&q^{-z}\nonumber\\
T_{\rm
eff}(q=0,\omega)/T&\sim&\left|\omega-\omega_0^{\rm{eff}}\right|^{-\sigma}.
\end{eqnarray}
For the particular case $c_a=0$ and $u_a=0$, $\sigma=1$, while
otherwise $\sigma\simeq 1-\epsilon/5$ to first order in
$\epsilon$\footnote{The divergence of the effective temperature for
small $q$ follows from simple scaling arguments. The divergence for
small $\omega-\omega^{\rm eff}_0$, however, results from the
non-trivial renormalization of $\omega_0$ as described by Eq.
(\ref{EffCorrRespOm}) and (\ref{ReOm0}).}. This singular behavior of
the effective temperature implies a violent breaking of the FD
relation. This is consistent with the fact that spontaneously
oscillating systems operate far from thermodynamic equilibrium.

\section{\label{Se.SumConcl}Summary and conclusion}

We have studied the critical behaviors of a large number of locally
coupled oscillators when approaching a homogeneous synchronization
transition from the disordered phase in a $d=4-\epsilon$ dimensional
space. On large length and time scales, the critical behaviors can
be described by a statistical field theory that is given by the
complex Ginzburg-Landau equation with additional noise and forcing
terms. At the critical point of a homogeneous oscillatory
instability, time-translational invariance is spontaneously broken
in the system. The field variable $Z$ in the complex Ginzburg-Landau
field theory is constructed in such a way that time-translations
correspond to global phase changes of this complex variable $Z$.
Within this framework, breaking of time-translational invariance
becomes formally similar to the traditional spontaneous symmetry
breaking known for other second-order phase transitions.

We have established the structure of the associated dynamic RG
within Wilson and Callan-Symanzik schemes, and performed the
calculations to two-loop order in perturbation theory. We have shown
that the critical point is formally related to the equilibrium phase
transition in the real Ginzburg-Landau $O(2)$ dynamic model A.
However, the RG flow of critical oscillators is defined in a larger
parameter space of non-equilibrium field theories and leads to a
renormalization of oscillation frequency and phase. The FD relation
is broken in the system, which can be characterized by an effective
frequency-dependent temperature, diverging at the effective
oscillation frequency with an anomalous power-law.

The formal analogy with an $O(2)$ symmetric dynamic field theory,
valid at the critical point, leads to several interesting results.
For $d>2$, the collective dynamics of coupled oscillators exhibits a
second order non-equilibrium phase transition. This phase transition
is a generalization of Hopf bifurcations, which are conventionally
defined in the context of nonlinear dynamics, to non-equilibrium
statistical physics. On the oscillating side of the bifurcation and
in the thermodynamic limit, the system exhibits long-range phase
order and coherent oscillations. In mean field theory, the universal
properties of this oscillating instability are captured by the
normal form known from nonlinear dynamics. Below the upper critical
dimension however, fluctuations become relevant and anomalous
scaling laws and critical exponents appear.

The case of critical coupled oscillators studied here provides a
further example for the emergence of an equilibrium universality
class in a non-equilibrium dynamic field theory. In non-equilibrium
systems with non-conserved order parameter, detailed balance is
often effectively restored at criticality \cite{taub02}. This is the
case e.g. for the model A dynamics of the real Ginzburg-Landau
theory with $Z_2$ symmetry \cite{haak84,grin85}, even when the
symmetry is broken by the non-equilibrium perturbations
\cite{bass94}, and for some of its generalizations to the $O(n)$
symmetry \cite{taub97}. In the present case, the detailed balance
condition is not restored for the physical variables, but appears
only in the oscillating reference frame associated with the
effective frequency and phase of the oscillations at the transition.
This emergence of detailed-balance symmetry at criticality imposes a
generic relation between the correlation and response functions of
coupled oscillators as given by Eq. (\ref{nFD}).

The structure of the RG flow studied here is singular at the upper
critical dimension $d_c=4$. Indeed, as depicted in Fig.
\ref{Fig.FlowDiagCS}, the line of Gaussian fixed points, which is
stable above $d=4$, becomes unstable for $d<4$ where only one
isolated stable fixed point remains. Our results obtained in an
epsilon expansion are valid close to the upper critical dimension
for $d=4-\epsilon$. We can speculate how our results are modified in
lower dimensions $d$. In analogy with the equilibrium $O(2)$ dynamic
model, we expect the phase order of the oscillations to vanish for
$d<2$, and to be quasi-long range exactly at the lower critical
dimension $d=2$. In the last case, spectral peaks on the oscillating
side of the Hopf bifurcation are expected to exhibit power-law tails
with non-universal exponents. If the formal analogy with the
equilibrium critical point found here in $d=4-\epsilon$ persists in
$d=2$, we would expect to see features of the Kosterlitz-Thouless
universality class \cite{kost73} in systems of coupled oscillators
in this dimension.

The different values of the space dimension $d$ of coupled
oscillators can be related to different realizations of coupled
nonlinear oscillators in various physical and biological systems.
The mean field limit $d>4$ is found in systems where oscillators are
coupled by long-range interactions. Examples for such a situation
are sarcomere oscillations in muscles. There, large numbers of
myosin motor proteins generate oscillations when interacting with
actin filaments, which represent tracks along which the motor
proteins move. Oscillations occur if the motor collection acts
against elastic elements and in the presence of a chemical fuel that
supplies the necessary energy. For stiff filaments, this situation
is well described by globally coupled motor proteins for which mean
field theory applies at the Hopf bifurcation \cite{juli97b}.

Systems of coupled oscillators in three dimensions could be realized
in oscillatory chemical processes in bulk solution. This is the case
e.g. of the Belousov-Zhabotinsky reaction which can be studied in
the framework of the complex Ginzburg-Landau equation (see e.g.
\cite{kapr95}). On mesoscopic scales, the system can be viewed as a
collection of interacting volume elements, each representing an
individual chemical oscillator. There, oscillations are subject to
fluctuations due to the finite number of reacting molecules present
in each volume element.

Coupled oscillators in two dimensions can be realized by oscillators
arranged on a surface. Such a situation may occur in the
electrosensory organ of some fish species where many electrically
oscillating cells constitute the sensory epithelium
\cite{brau94,neim01}. Critical oscillators coupled in two dimensions
can also in principle be realized in artificial systems.
Nanotechnology aims to build functional units on the sub-micrometer
scale. Large arrays of nano-oscillators on patterned substrates
coupled to their neighbors by elastic or viscous effects would
provide a 2-dimensional realization of our field theory. Finally,
the case $d=0$ corresponds to a single noisy oscillator. Here,
fluctuations destroy the Hopf bifurcation and only its signatures
can be observed. In the context of biological systems, an example is
the spontaneous oscillations of the mechano-sensory organelle of
auditory hair cells \cite{mart01b,ospe01}. Here, the critical
divergence of the linear response function is ideally suited for
signal detection.

In order to observe the critical exponents discussed here,
homogeneous chemical oscillations in a bulk system with $d=3$ (which
corresponds here to $\epsilon=1$) would be a good candidate.
However, the critical exponents attached to the RG fixed point are
only observable when the system is observed sufficiently close to
the critical point. The range and experimental accessibility of this
critical regime can be estimated by a Ginzburg criterion, see
Appendix \ref{ginzburg}. Assuming that in a chemical system, the
Hopf bifurcation occurs if a molecular concentration $\rho$ exceeds
a critical value $\rho_c$, the critical regime corresponds to
\begin{equation}
\frac{\vert \rho-\rho_c\vert}{\rho_c}< \frac{\omega_0}{\rho^2
\tau_c^2 c^{3}},\label{ginz1}
\end{equation}
where $\omega_0$ is the oscillation frequency and $\tau_c$ denotes a
chemical reaction time. The coefficient $c$ here is the bare
coefficient describing the coupling of oscillators in the complex
Ginzburg-Landau equation. We can rewrite this expression as
\begin{equation}
\frac{\vert \rho-\rho_c\vert}{\rho_c}< \frac{\omega_0\tau_c}{(\rho
\,l_c^3)^2}\left(\frac{D_m}{c}\right)^3,\label{ginz2}
\end{equation}
where $l_c^2=D_m\tau_c$ is a reaction length and $D_m$ denotes a
microscopic diffusion coefficient. Since $\omega_0\tau_c\ll1$
(oscillations are slow compared to fast reaction times) and
$l_c^3\rho\gg 1$ (the volume per molecule is small compared to the
reaction volume), accessibility to the critical regime requires that
$D_m\gg c$. This condition is satisfies if $c$ becomes small. This
happens in particular if a Turing instability is approached. At the
point where such an instability occurs, the coefficient $c$ changes
sign and stationary spatial patterns appear. Our analysis suggests
that before such a point is reached, the critical regime of the Hopf
bifurcation becomes accessible. Therefore, the scaling behaviors and
critical exponents discussed here could be experimentally observable
in oscillating chemical systems.

\section*{\label{Se.Ackn}Acknowledgments}

We thank Edouard Br\'ezin, Erwin Frey and Kay Wiese for useful
discussions.

\appendix

\section{\label{Ap.GraphRepr}Feynman diagrams of the
perturbation theory}

\begin{figure}[h]
\scalebox{0.4}{
\includegraphics{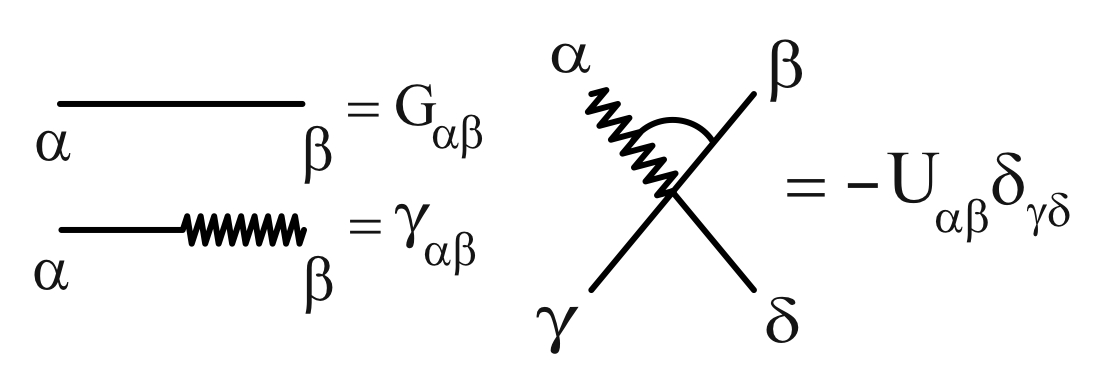}
}
\caption[fig4]{\label{FeynElem} Graphic representation of the
propagators $G_{\alpha\beta}$ and $\gamma_{\alpha\beta}$, and of the
vertex $U_{\alpha\beta}\delta_{\gamma\delta}$.}
\end{figure}

Here, we present the graphic representation of the terms of the
expansion series that we used for the calculations to one and
two-loop orders in perturbation theory. The expansion series of the
action of the complex Ginzburg-Landau theory given by Eq.
(\ref{ActPhi}) can be represented by Feynman diagrams as usual. The
free propagator of the theory, calculated from the Gaussian part of
the action $\mathcal{S}_{\textrm{R}}$ of the decomposition
(\ref{ActDecCS}) in a Callan-Symanzik RG scheme, is given by
\begin{eqnarray}
G_{\alpha\beta}^0(\mathbf{q},\omega)&=&\frac{2D}{{\left| \Delta
\right|}^2} \left( \begin{array}{cc}
\omega^{2} + (c_a\mathbf{q}^2)^2 + (c\mathbf{q}^2)^{2} & 2i\omega c_a\mathbf{q}^2\\
-2i\omega c_a\mathbf{q}^2 & \omega^{2} + (c_a\mathbf{q}^2)^2 +
(c\mathbf{q}^2)^{2}
\end{array} \right)\nonumber\\
\nonumber\\
\gamma_{\alpha\beta}^0(\mathbf{q},\omega)&=&\frac{1}{\Delta} \left(
\begin{array}{cc}
-i\omega+c\mathbf{q}^2 & c_a\mathbf{q}^2\\
-c_a\mathbf{q}^2 & -i\omega+c\mathbf{q}^2
\end{array} \right),
\end{eqnarray}
where $\Delta=(-i\omega+c\mathbf{q}^2)^2+(c_a\mathbf{q}^2)^2$. The
interaction vertex reads:
\begin{equation}
-U_{\alpha\beta}\delta_{\gamma\delta}=-\mu^{\epsilon}(4\pi)^{-\epsilon/2}g_{\alpha\beta}\delta_{\gamma\delta},
\end{equation}
where $g_{\alpha\beta}$ is given by Eq. (\ref{CouplConstGMat}).
Graphic representations of these elements are displayed in Fig.
\ref{FeynElem}.
Due to the presence of the non-diagonal element
$u_a$ in the matrix $U_{\alpha\beta}$, the interaction vertex
contains three non-equivalent types of external ``legs''. The symbol
used for the interaction vertex in Fig. \ref{FeynElem} indicates
this fact. The expressions of the counter-terms of the theory are
directly visible on the decomposed expressions (\ref{DiffActCS2})
and (\ref{DiffActCS3}) of the action of the theory.

\begin{figure}[h]
\scalebox{0.4}{
\includegraphics{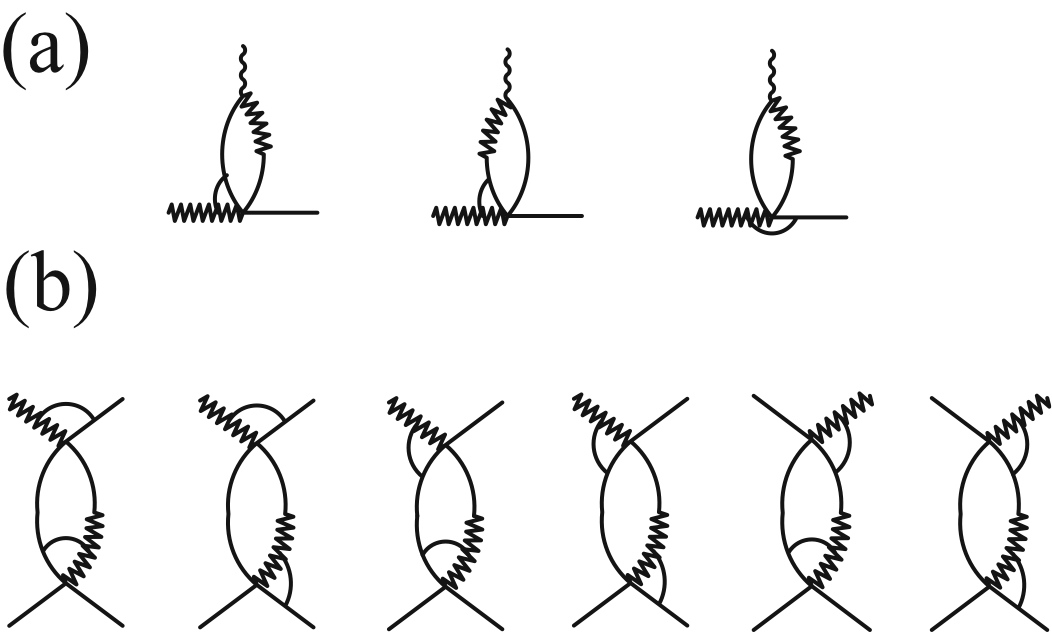}
}
\caption[fig5]{\label{diag1B} Feynman diagrams contributing to the
renormalization of the theory to one-loop order. (a) Renormalization
of the ``mass operator'' $[\tilde{\phi}_{\alpha}\phi_{\beta}]$. (b)
Renormalization of the interaction vertex.}
\end{figure}

\begin{figure}[h]
\scalebox{0.4}{
\includegraphics{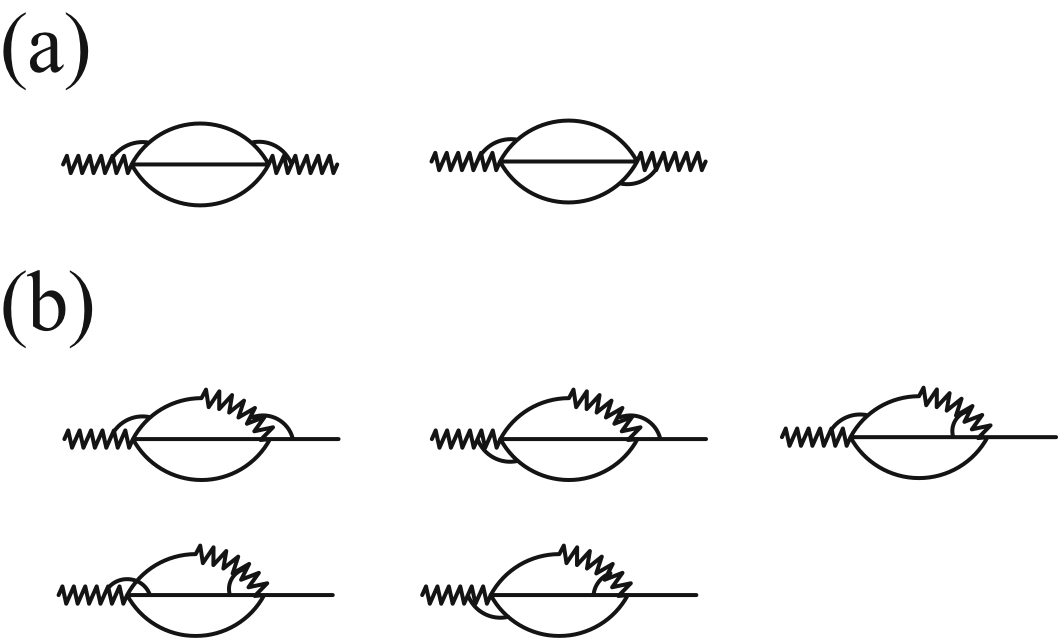}
}
\caption[fig6]{\label{diag2B} Feynman diagrams contributing to the
renormalization of the propagator to two-loop order in perturbation
theory. (a) Contribution to the renormalization of the noise
amplitude parameter $D(Z_{\omega}Z_{\tilde{\phi}}-1)$ in the
counter-terms $\delta\mathcal{S}$ in Eq. (\ref{DiffActCS2}). (b)
Contribution to the renormalization of the other counter-terms to
the propagator.}
\end{figure}

Fig. \ref{diag1B} displays the diagrams that contribute to the
renormalization of the ``mass operator''
$[\tilde{\phi}_{\alpha}\phi_{\beta}]$ and the vertex to one-loop
order, and Fig. \ref{diag2B} shows the diagrams that contribute to
the renormalization of the propagator to two-loop order in
perturbation theory.

\section{\label{Ap.ExplExpr}Explicit expressions of the calculated perturbation series}

We present here explicit expressions of the Feynman integrals
associated with the diagrams displayed previously, to one and
two-loop order in perturbation theory. Using the notation introduced
in Subsection \ref{Sse.GenRGForm}, we denote by
$\Gamma^{(\tilde{N},N,L)(n)}_{\{\alpha_i\},\{\beta_j\}}$ the
contribution of the $n$-loop order to the vertex function with
$\tilde{N}$ and $N$ truncated external legs corresponding
respectively to the fields $\tilde{\phi}_{\alpha_i}$ and
$\phi_{\beta_j}$, and with $L$ insertions of the mass operator
$[\tilde{\phi}_{\alpha}\phi_{\beta}]$.

\subsection{\label{ExplExpr1Loop} One-loop order}

To one-loop order in perturbation theory, the propagator of the
theory is not renormalized. The renormalization of the ``mass
operator'' $[\tilde{\phi}_{\alpha}\phi_{\beta}]$ is given by
\begin{equation}
\Gamma^{(1,1,1)(1)}_{\alpha,\beta}=r\mu^{\epsilon}(4\pi)^{-\epsilon/2}\,J_d
\,\left[g_{\beta\gamma}\delta_{\alpha\sigma}+
g_{\beta\sigma}\delta_{\alpha\gamma}+g_{\beta\alpha}\delta_{\gamma\sigma}\right]
\,\delta_{\gamma\sigma},
\end{equation}
where $g_{\alpha\beta}$ is given by Eq. (\ref{CouplConstGMat}). The
expression of the integral $J_d$ is given below. Because of phase
invariance symmetry, described for
$\Gamma^{(1,1,1)(1)}_{\alpha,\beta}$ by Eq. (\ref{RotSym}), only two
terms need to be calculated.

The renormalization of the vertex function to one-loop order in
perturbation theory at a symmetry point of the configuration of
external momenta, respects the following symmetry:
\begin{equation}\label{Gam4Sym}
\Gamma^{(1,3,0)(1)}_{\alpha_1,\beta_1\beta_2\beta_3}(k_1,k_2,k_2,k_2)=\left[
M_{\alpha_1\beta_1}\delta_{\beta_2\beta_3}+M_{\alpha_1\beta_2}\delta_{\beta_1\beta_3}+M_{\alpha_1\beta_3}\delta_{\beta_1\beta_2}\right]
\times (2\pi)^{d+1}\delta^{d+1}(k_1+3k_2),
\end{equation}
where $M_{\alpha\beta}$ has the phase invariance symmetry
(\ref{RotSym}). Therefore, the renormalized interaction vertex of
the theory remains of the same structure as the original one, and
only two independent terms need to be calculated, e.g.
$\Gamma^{(1,3,0)(1)}_{1,111}$ and $\Gamma^{(1,3,0)(1)}_{2,111}$. For
vanishing external frequencies, we get
\begin{eqnarray}
\Gamma^{(1,3,0)(1)}_{1,111}&=&24\mu^{2\epsilon}(4\pi)^{-\epsilon}\left[
g^2\left(\frac{1}{2}I_d+2J_d\right)+g_a^2\left(-\frac{1}{2}I_d\right)+2gg_a\left(\frac{i}{2}I_d+iJ_d\right)+\textrm{c.c.}\right]\nonumber\\
\Gamma^{(1,3,0)(1)}_{2,111}&=&24\mu^{2\epsilon}(4\pi)^{-\epsilon}\left[
g^2\left(-\frac{i}{2}I_d\right)+g_a^2\left(\frac{i}{2}I_d+2iJ_d\right)+2gg_a\left(\frac{1}{2}I_d+J_d\right)+\textrm{c.c.}\right],\nonumber\\
\end{eqnarray}
where ``c.c.'' denotes the complex conjugated value.

In the previous expressions, the integrals $I_d$ and $J_d$ are given
by
\begin{eqnarray}
I_d(c,c_a,\mathbf{q})&=&\int_{\mathbf{p}}\frac{D}{c\mathbf{p}^2(c+ic_a)\left[\mathbf{p}^2+(\mathbf{p}-\mathbf{q})^2\right]}\nonumber\\
J_d(c,c_a,\mathbf{q})&=&\int_{\mathbf{p}}\frac{D}{c\mathbf{p}^2\left[(c-ic_a)\mathbf{p}^2+(c+ic_a)(\mathbf{p}-\mathbf{q})^2\right]}.
\end{eqnarray}
Within a dimensional regularization scheme, they read:
\begin{eqnarray}\label{1LoopInt}
I_d(c,c_a,\mathbf{q})&=&\frac{D}{c(c+ic_a)}\frac{q^{d-4}}{(4\pi)^{d/2}}\Gamma\left(\frac{4-d}{2}\right)
\int_0^1\frac{x^{\frac{d-4}{2}}}{(1+x)^{d-2}}\,d x\nonumber\\
&=&\frac{D}{c(c+ic_a)}\frac{1}{(4\pi)^2(4\pi)^{-\epsilon/2}}\frac{1}{\epsilon}\big(1+\mathcal{O}(\epsilon)\big)\nonumber\\
\nonumber\\
J_d(c,c_a,\mathbf{q})&=&\frac{D}{c}\frac{q^{d-4}}{(4\pi)^{d/2}}\Gamma\left(\frac{d}{2}\right)\Gamma\left(\frac{4-d}{2}\right)\nonumber\\
&&\times\int_0^1\frac{d
x}{(1-x+2cx)^2}\left[x(c+ic_a)\frac{1-x+x(c-ic_a)}{(1-x+2cx)^2}\right]^{\frac{d-4}{2}}\nonumber\\
&=&\frac{D}{c^2}\frac{1}{(4\pi)^2(4\pi)^{-\epsilon/2}}\frac{1}{\epsilon}\big(1+\mathcal{O}(\epsilon)\big),
\end{eqnarray}
where $q=|\mathbf{q}|$.

\subsection{\label{ExplExpr2Loop}Explicit expressions of the Feynman
integrals to two-loop order}

Since we are looking for the first non-trivial corrections to the
critical behaviors in perturbation theory, we only need here to
renormalize the propagator, which to one-loop order was not
renormalized. Following the same notations as previously, and for
vanishing external frequencies, we have
\begin{eqnarray}
\Gamma^{(2,0,0)(2)}_{\alpha_1\alpha_2}(\mathbf{k})&=&\mu^{2\epsilon}(4\pi)^{-\epsilon}(g^2+g_a^2)\left[
2I_A+6I_B\right]\delta_{\alpha\beta}\nonumber\\
\Gamma^{(1,1,0)(2)}_{1,1}(\mathbf{k})&=&2\mu^{2\epsilon}(4\pi)^{-\epsilon}\left[
g^2(4J_A+7J_C+J_E)+g_a^2(4J_A-7J_C-J_E)+2gg_a(7J_D-J_F)\right]\nonumber\\
\Gamma^{(1,1,0)(1)}_{2,1}(\mathbf{k},0)&=&2\mu^{2\epsilon}(4\pi)^{-\epsilon}\left[g^2(-4J_B-7J_D+J_F)+g_a^2(-4J_B+7J_D-J_F)+2gg_a(7J_C+J_E)\right].
\end{eqnarray}
Here
\begin{eqnarray}
I_A&=&\frac{1}{2}(I_{d,\epsilon'\epsilon''}(c_a)+I_{d,\epsilon'\epsilon''}(-c_a)) \quad \epsilon'=+1 \quad \epsilon''=-1\nonumber\\
I_B&=&\frac{1}{2}(I_{d,\epsilon'\epsilon''}(c_a)+I_{d,\epsilon'\epsilon''}(-c_a)) \quad \epsilon'=-1 \quad \epsilon''=+1\nonumber\\
J_A&=&\frac{1}{2}(J_{d,\epsilon'\epsilon''}(c_a)+J_{d,\epsilon'\epsilon''}(-c_a))
\quad \epsilon'=+1 \quad \epsilon''=-1\nonumber\\
J_B&=&\frac{i}{2}(J_{d,\epsilon'\epsilon''}(c_a)-J_{d,\epsilon'\epsilon''}(-c_a))
\quad \epsilon'=+1 \quad \epsilon''=-1\nonumber\\
J_C&=&\frac{1}{2}(J_{d,\epsilon'\epsilon''}(c_a)+J_{d,\epsilon'\epsilon''}(-c_a))
\quad \epsilon'=-1 \quad \epsilon''=+1\nonumber\\
J_D&=&\frac{i}{2}(J_{d,\epsilon'\epsilon''}(c_a)-J_{d,\epsilon'\epsilon''}(-c_a))
\quad \epsilon'=-1 \quad \epsilon''=+1\nonumber\\
J_E&=&\frac{1}{2}(J_{d,\epsilon'\epsilon''}(c_a)+J_{d,\epsilon'\epsilon''}(-c_a))
\quad \epsilon'=-1 \quad \epsilon''=-1\nonumber\\
J_F&=&\frac{i}{2}(J_{d,\epsilon'\epsilon''}(c_a)-J_{d,\epsilon'\epsilon''}(-c_a))
\quad \epsilon'=-1 \quad \epsilon''=-1,
\end{eqnarray}
and
\begin{eqnarray}
I_{d,\epsilon'\epsilon''}(c,c_a,\mathbf{k})&=&\int_{\mathbf{p},\mathbf{q}}\frac{D^3}
{c\mathbf{q}^2c(\mathbf{p}-\mathbf{q})^2c(\mathbf{k}-\mathbf{p})^2\left[
(c+ic_a)\mathbf{q}^2+(c+i\epsilon'c_a)(\mathbf{p}-\mathbf{q})^2+(c+i\epsilon''c_a)(\mathbf{k}-\mathbf{p})^2\right]}\nonumber\\
J_{d,\epsilon'\epsilon''}(c,c_a,\mathbf{k},\omega_k)&=&\int_{\mathbf{p},\mathbf{q}}\frac{D^2}{c\mathbf{q}^2c(\mathbf{p}-\mathbf{q})^2
\left[(c+ic_a)\mathbf{q}^2+(c+i\epsilon'c_a)(\mathbf{p}-\mathbf{q})^2+(c+i\epsilon''c_a)(\mathbf{k}-\mathbf{p})^2+i\omega_k\right]}.\nonumber\\
\end{eqnarray}
The expressions of these integrals as a function of the space
dimension $d$ are too large to be displayed here. We therefore just
report the expressions of their divergent parts as $\epsilon$ goes
to zero:
\begin{eqnarray}
I_{d,\epsilon'\epsilon''}(c,c_a,\mathbf{k})
&=&\frac{D^3}{c^4}\frac{1}{(4\pi)^4(4\pi)^{-\epsilon}}\frac{1}{\epsilon}\,H_{\epsilon'\epsilon''}(\bar{c}_a)\,(1+\mathcal{O}(\epsilon))\nonumber\\
\frac{\partial
J_{d,\epsilon'\epsilon''}(c,c_a,\mathbf{k},\omega_k)}{\partial
c\mathbf{k}^2}
&=&-\frac{D^2}{c^4}\frac{1}{(4\pi)^4(4\pi)^{-\epsilon}}\frac{1}{\epsilon}\,
K_{1,\epsilon'\epsilon''}(\bar{c}_a)\,(1+\mathcal{O}(\epsilon))\nonumber\\
\frac{\partial
J_{d,\epsilon'\epsilon''}(c,c_a,\mathbf{k},\omega_k)}{\partial
i\omega_k}
&=&-\frac{D^2}{c^3}\frac{1}{(4\pi)^4(4\pi)^{-\epsilon}}\frac{1}{\epsilon}\,
K_{2,\epsilon'\epsilon''}(\bar{c}_a)\,(1+\mathcal{O}(\epsilon)),
\end{eqnarray}
where $\bar{c}_a=c_a/c$. The integrals
$H_{\epsilon'\epsilon''}(\bar{c}_a)$,
$K_{1,\epsilon'\epsilon''}(\bar{c}_a)$ and
$K_{2,\epsilon'\epsilon''}(\bar{c}_a)$ are given by expressions that
are similar to the ones found to the one-loop order (Eq.
(\ref{1LoopInt})), which for $d=4$ reduce to integrals of rational
fractions with complex parameters. The result of these integrations
read:
\begin{eqnarray}\label{Eq.IntExpr1}
K_{1,+1+1}(\bar{c}_a)&=&I_1(\bar{c}_a)=\frac{1}{6(1+i\bar{c}_a)}\nonumber\\
K_{1,-1-1}(\bar{c}_a)&=&I_2(\bar{c}_a)=\frac{2+i\bar{c}_a}{4(3+i\bar{c}_a)}\nonumber\\
K_{1,-1+1}(\bar{c}_a)&=&I_3(\bar{c}_a)=I_2(-\bar{c}_a)\nonumber\\
K_{1,+1-1}(\bar{c}_a)&=&I_4(\bar{c}_a)=\frac{1-i\bar{c}_a}{6-2i\bar{c}_a}\nonumber\\
K_{2,+1+1}(\bar{c}_a)&=&I_5(\bar{c}_a)=\frac{1}{(1+i\bar{c}_a)^2}\ln\left[\frac{4}{3}\right]\nonumber\\
K_{2,-1-1}(\bar{c}_a)&=&I_6(\bar{c}_a)=\frac{1}{(1-i\bar{c}_a)^2}\ln\left[\frac{4}{3+i\bar{c}_a}\right]\nonumber\\
K_{2,-1+1}(\bar{c}_a)&=&I_7(\bar{c}_a)=I_6(-\bar{c}_a)\nonumber\\
K_{2,+1-1}(\bar{c}_a)&=&I_8(\bar{c}_a)=\frac{1}{(1-i\bar{c}_a)^2}\ln\left[\frac{4}{(3-i\bar{c}_a)(1+i\bar{c}_a)}\right]\nonumber\\
H_{+1-1}(\bar{c}_a)&=&H_{-1+1}(\bar{c}_a)=I_9(\bar{c}_a),
\end{eqnarray}
where $I_9(\bar{c}_a)$ can be represented by
\begin{eqnarray}\label{I9}
I_9(\bar{c}_a)&=&\frac{1}{(1+\bar{c}_a^2)}\left\{
4\ln[2]-i\arctan\left[\frac{2\bar{c}_a}{3+{\bar{c}_a}^2}\right]
-\frac{1}{2}\ln\left[9+10\bar{c}_a^2+\bar{c}_a^4\right]\right.\nonumber\\
&&\left.-\ln\left[-i\left(1+2i\bar{c}_a+\bar{c}_a^2\right)\right]+
\ln\left[-i\frac{1+2i\bar{c}_a+\bar{c}_a^2}{3+2i\bar{c}_a+\bar{c}_a^2}\right]\right\}\nonumber\\
&&-\frac{1}{\sqrt{(1+i\bar{c}_a)^2}}
\left\{\ln{\left[-i\frac{\sqrt{(1+i\bar{c}_a)^2}(1-i\bar{c}_a)}{\left(1+i\bar{c}_a+\sqrt{(1+i\bar{c}_a)^2}\right)^2}\right]}
-\ln{\left[-i\frac{2+i\bar{c}_a+\bar{c}_a^2-\sqrt{(1+i\bar{c}_a)^2}}{2+i\bar{c}_a+\bar{c}_a^2+\sqrt{(1+i\bar{c}_a)^2}}\right]}\right\}.\nonumber\\
\end{eqnarray}
Because of the ambiguity in the definition of the complex logarithm,
the integrals $I_6(\bar{c}_a)$, $I_7(\bar{c}_a)$, $I_8(\bar{c}_a)$
and $I_9(\bar{c}_a)$ are not uniquely defined by these
expressions\footnote{Note that, because of this ambiguity, the
expression (\ref{I9}) can not be simplified further.}. To specify
entirely these functions, one needs to use their values for
$\bar{c}_a=0$, which read:
\begin{eqnarray}\label{IntExprCa=0}
I_6(\bar{c}_a=0)&=&I_7(\bar{c}_a=0)=I_8(\bar{c}_a=0)=\ln\left[\frac{4}{3}\right]\nonumber\\
I_9(\bar{c}_a=0)&=&3\ln\left[\frac{4}{3}\right].
\end{eqnarray}
The entire functions are then defined as the unique analytic
prolongations of the expressions (\ref{Eq.IntExpr1}) and (\ref{I9}),
defined in the vicinity of $\bar{c}_a=0$ together with the
specification (\ref{IntExprCa=0}).

\section{\label{Ap.ExprRGEq}Explicit expressions of the renormalization group
equations}

With units such that $c=1$, the function $\beta_c(g,g_a,c_a)$ is
given by:
\begin{eqnarray}\label{beta_c}
\beta_c(g,g_a,c_a)&=&\mathcal{D}^2\Big[(g^2+g_a^2)\textrm{Im}(-I_4)+(g^2-g_a^2)\textrm{Im}(2I_2)+2gg_a\textrm{Re}(-2I_2)\Big]\nonumber\\
&&+\mathcal{D}^2\bar{c}_a\Big[(g^2+g_a^2)\textrm{Re}(I_4)+(g^2-g_a^2)\textrm{Re}(2I_2)+2gg_a\textrm{Im}(2I_2)\Big]\nonumber\\
&&+\mathcal{D}^2c_a(c_a+c_a^{-1})\Big[(g^2+g_a^2)\textrm{Im}(I_8)+(g^2-g_a^2)\textrm{Im}(-2I_6)+2gg_a\textrm{Re}(2I_6)\Big],\nonumber\\
\end{eqnarray}
where $\mathcal{D}=4D/(4\pi)^2$\footnote{Note that the integral
$I_9$ does not enter the expression of the beta function here.
Therefore, this quantity is not affecting the determination of the
fixed points to this order in perturbation theory.}. The function
$\rho_c(c_a)$, which is defined in Eq. (\ref{Rho_c}) and determines
the fixed points of the theory to too-loop order, is given by:
\begin{eqnarray}\label{rho_c}
\rho_c(c_a)&=&-\frac{1}{50}(1+c_a^2)\left\{4(1-c_a^2)\textrm{Im}\left[\frac{\ln\left(\frac{4}{3+i
c_a}\right)}{(1-i
c_a)^2}\right]-2(1+c_a^2)\textrm{Im}\left[\frac{\ln\left(\frac{4}{3+2
i c_a+c_a^2}\right)}{(1-i c_a)^2}\right]-8 c_a\textrm{Re}\left[\frac{\ln\left(\frac{4}{3+i
c_a}\right)}{(1-i c_a)^2}\right]\right\},\nonumber\\
\end{eqnarray}
where the three complex logarithms are defined by prolongation of
their real values for $c_a=0$. This function is displayed in Fig.
(\ref{Fig.RhoCa}).

Finally, the other Wilson's functions of the theory are given by the
following expressions:
\begin{equation}
\gamma_{\theta}(g,g_a,c_a)=\mathcal{D}^2\Big[(g^2+g_a^2)\textrm{Im}(I_8)+(g^2-g_a^2)\textrm{Im}(-2I_6)
+2gg_a\textrm{Re}(2I_6)\Big],
\end{equation}
and:
\begin{eqnarray}
\gamma(g,g_a,c_a)&=&\mathcal{D}^2\Big[(g^2+g_a^2)\textrm{Re}(3I_4-I_8-I_9)+(g^2-g_a^2)\textrm{Re}(6I_2-2I_6)
+2gg_a\textrm{Im}(6I_2-2I_6)\Big]\nonumber\\
&&+3\mathcal{D}^2 c_a\Big[(g^2+g_a^2)\textrm{Im}(I_8)+(g^2-g_a^2)\textrm{Im}(-2I_6)+2gg_a\textrm{Re}(2I_6)\Big]\nonumber\\
\nonumber\\
\gamma_{\omega}(g,g_a,c_a)&=&\mathcal{D}^2\Big[(g^2+g_a^2)\textrm{Re}(-I_4+I_8)+(g^2-g_a^2)\textrm{Re}(-2I_2+2I_6)
+2gg_a\textrm{Im}(-2I_2+2I_6)\Big]\nonumber\\
&&-\mathcal{D}^2 c_a\Big[(g^2+g_a^2)\textrm{Im}(I_8)+(g^2-g_a^2)\textrm{Im}(-2I_6)+2gg_a\textrm{Re}(2I_6)\Big]\nonumber\\
\nonumber\\
\tilde{\gamma}(g,g_a,c_a)&=&\mathcal{D}^2\Big[(g^2+g_a^2)\textrm{Re}(I_4-I_8+I_9)+(g^2-g_a^2)\textrm{Re}(2I_2-2I_6)
+2gg_a\textrm{Im}(2I_2-2I_6)\Big]\nonumber\\
&&+\mathcal{D}^2
c_a\Big[(g^2+g_a^2)\textrm{Im}(I_8)+(g^2-g_a^2)\textrm{Im}(-2I_6)+2gg_a\textrm{Re}(2I_6)\Big].
\end{eqnarray}

\section{Ginzburg criterion}
\label{ginzburg} The critical behaviors described by the RG fixed
point are valid in the proximity of the critical point. In practice,
one can estimate how close to the critical point an experiment needs
to be performed in order for the critical behaviors to become
observable. Further away from the critical point, when mean field
theory is still an appropriate approximation, nontrivial critical
exponents are unobservable. The Ginzburg criterion estimates the
breakdown of mean field theory at the point where the variance of
order parameter fluctuations exceeds its average \cite{ginz60}. For
simplicity, we discuss here the real equation (\ref{cgle}) with
$u_a=0$ and $c_a=0$. Based on this criterion, mean field theory is
valid if
\begin{equation}\label{ginz}
\frac{D\xi^{-(d-2)}}{c}<\frac{\vert r\vert}{u},
\end{equation}
with $\xi^2=c/\vert r\vert$. We now relate this expression to a bulk
chemical system where oscillations occur if the concentration $\rho$
of some species exceeds a critical value $\rho_c$: $r\simeq a
(\rho_c-\rho)$ where $a$ is a proportionality coefficient. In this
case, the real part of $Z$ is related to the concentration
fluctuations $\delta \rho/\rho$. As a consequence, $r$ and $u$ have
dimensions of an inverse time, while the noise strength $D$ is a
volume per unit of time. The coefficient $c$ has units of a
diffusion coefficient. We can estimate the fluctuations of the
number $N$ of molecules in a reference volume $v$ after a time
$\tau$ using $\partial_t \delta\rho\simeq \rho\eta$ as:
\begin{equation}
\langle\delta N^2\rangle\simeq \int_v d^d x\, d^d x' \langle\delta
\rho({\bf x})\delta\rho({\bf x}')\rangle \simeq \rho^2 D v \tau.
\end{equation}
We estimate $\langle\delta N^2\rangle/N^2\simeq 1/N\simeq
1/l_c^3\rho$ within a reaction volume $v=l_c^3$ and a reaction time
$\tau_c$, with $l_c^2=D_m \tau_c$ and where $D_m$ is a microscopic
diffusion coefficient. From this, it follows that $D\simeq 1/(\rho
\tau_c)$. For $d=3$, Eq. (\ref{ginz}) implies that mean field theory
is valid if
\begin{equation}
\frac{\vert \rho-\rho_c\vert}{\rho_c} > \frac{u^2 D^2}{r_0 c^3},
\end{equation}
where $r_0=a\rho_c$. Estimating $r_0\sim u\sim \omega_0$, we find
Eq. (\ref{ginz1}) and (\ref{ginz2}).

\end{document}